\documentclass[]{article}
\usepackage{amssymb,epsf,rotate,amsbsy}
\newcommand{\e}{\varepsilon_6}

\newcommand{\kdp}{KH$_2$PO$_4$}
\newcommand{\dekdp}{KD$_2$PO$_4$}

\newcommand{\be}{\begin{equation}}
\newcommand{\mframe}{}
\newcommand{\ee}{\end{equation}}
\newcommand{\bea}{\begin{eqnarray}}
\newcommand{\eea}{\end{eqnarray}}
\newcommand{\s}{\sigma_6}
\newcommand{\eps}{\varepsilon}

\begin{document}

\title{LONGITUDINAL FIELD INFLUENCE ON THE PHASE TRANSITION AND
PHYSICAL PROPERTIES OF THE KH$_2$PO$_4$ FAMILY FERROELECTRICS}

\author{R.R.Levitskii, A.P.Moina, B.M.Lisnii\\
\it Institute for Condensed Matter Physics \\
\it 1 Svientsitskii St., 79011 Lviv, Ukraine\\
\tt email: alla@icmp.lviv.ua}

\date{}
\maketitle
\renewcommand{\theequation}{\arabic{section}.\arabic{equation}}

\sloppy

\begin{abstract}
We verify whether the previously developed model of a
KD$_2$PO$_4$ crystal, with the shear strain $\varepsilon_6$ taken into
account, is able to describe the  longitudinal
electric field $E_3$ influence  on
the KH$_2$PO$_4$ family ferroelectrics. Major effects of  the
strain $\e$ are splitting of the Slater energies of the
short-range correlations
and the effective field created by piezoelectric coupling. Calculated
$T_{\rm C}-E_3$ phase diagrams, field dependences of
polarization, susceptibility, and elastic constant of
deuterated KD$_2$PO$_4$ and pure KH$_2$PO$_4$ well accord with the
available experimental data. For a consistent description of all dielectric
and piezoelectric characteristics of the crystals, phonon degrees of
freedom must be taken into account.

\end{abstract}

\section{Introduction}
\normalsize
Studies of ferroelectric crystals behavior under different external factors
such as
pressure or electric field provide an important information about the
phase transition mechanisms in these crystals as well as of their
dielectric and piezoelectric responses.  Thus, studies of the
transition temperature dependence on the geometry of a hydrogen bond
revealed a universal dependence of  the transition  temperature on the
distance between equilibrium proton (deuteron) positions on a bond in
several crystals of the \kdp\ family   \cite{Nel5,our!}.
 Studies of the effects produced by external
fields conjugate to the order parameter (shear stress $\s$ and electric
field $E_3$ in these crystals) permit to elucidate the role of structural
changes occurring at the phase transition, such as appearance of the shear
strain $\e$.

In our recent paper \cite{our!!} we modified the proton ordering model
for deuterated \dekdp\ type crystals so that the shear strain $\e$ was
taken into account.  We obtained a fair
agreement of the theoretical results for associated with strain $\e$
dielectric, elastic, and piezoelectric characteristics of \dekdp\ (piezoelectric
constants $d_{36}$, $e_{36}$, $h_{36}$, $g_{36}$, elastic constants
 $c^P_{66}$, $c^E_{66}$ and compliances $s^P_{66}$, $s^E_{66}$,
and dielectric permittivity $\eps_{33}$) with
the relevant experimental data.

A consistent taking into account of the
changes caused by the strain $\e$ in the system symmetry  revealed that
this strain splits the energies of short-range deuteron configurations.
Thus, the lowest level of the so-called up and down configurations and
the energies of lateral and single-ionized configurations are split. Splitting
of up/down and ionized configurations gives rise to single- and three-particle
spin terms in the short-range Hamiltonian. Our numerical calculations
have shown that by fitting to experimental data one is not able to determine
the magnitude of the splitting of up/down configurations $\delta_{s6}$, since
if a certain ratio between $\delta_{s6}$ and the magnitude of piezoelectric
internal field $\psi_6$ \cite{stas1}, the experiment is well described in a rather
wide range of $\delta_{s6}$.  A good agreement with the experimental
data for temperature behavior of the above mentioned physical characteristics
of \dekdp\ is also obtained in a simplified version of the model
\cite{jps3},
where we neglected splitting of the up/down and
ionized configurations and took into account only splitting of lateral configurations
and piezoelectric internal fields.

In \cite{our!!} we show that the transition temperature
of \dekdp\ type crystals is increased by the stress $\s$, and
the jumps of the order parameter are decreased. At the
constructed phase diagram there are two critical points where
the phase equilibrium curves terminate. Stresses above critical smear
off the phase transition  and lead to smooth temperature dependences
of polarization and strain. Correspondingly, the peak values of
those characteristics of the crystal that have peculiarities
at the transition points (the longitudinal dielectric permittivity,
compliance $s_{66}^E$, piezomodules $d_{36}$ and $e_{36}$, and specific
heat) increase with stress.  These peaks are the highest at
the critical stress, whereas at stresses above critical their temperature
curves  are smooth. Such a behavior is peculiar to the ferroelectrics
with  the first order phase transition in external fields conjugate to the
order parameter.

Studies of the stress $\s$ influence on the \kdp\ family crystals,
though interesting they are, remain quite abstract, since their experimental
verification appears to be very difficult. Instead, measurements in electric
fields are a more accessible experimental technique.
The longitudinal electric field $E_3$ should be considered, since
its influence on the physics of the phase transition the relevant
characteristics of the \kdp\ type crystals is analogous to the influence
of the shear stress $\s$ and is well studied experimentally
\cite{Schmidt,Litov,Chabin,Sidnenko,Sidnenko2,Kobayashi,l366,sugie}.

In this paper we shall show that the developed theory is capable,
without introducing any additional fitting parameter, of describing
the behavior of the phase transition, polarization, and other
characteristics of a deuterated \dekdp\ in electric field
$E_3$, and attempt to apply this theory also for the description
of the relevant characteristics of a pure \kdp, with tunneling neglected.

\section{Phenomenological analysis}
Strictly speaking, for the \kdp\ type crystals a expansion of thermodynamic
potential $G$  in polarization $P_3$ and strain $\eps_6$
\begin{eqnarray}
\label{landau1}
&&G=G_0+\frac12\left(a_1P_3^2+a_2P_3\eps_6+a_3\eps_6^2\right)+
      \frac14\left(b_1P_3^4+b_2P_3^3\eps_6+b_3P_3^2\eps_6^2\right)+
      {}\nonumber\\
&&\quad{}+
\frac16\left(c_1P_3^6+c_2P_3^5\eps_6+c_3P_3^4\eps_6^2\right)-
  \bar vE_3P_3-\bar v\sigma_6\eps_6  \end{eqnarray}
 is the most
consistent. It allows one to perform the analysis for the both
fields conjugate to the order parameter -- electric field $E_3$
and shear stress $\s$. Then the relevant order parameter, after which the phase
transition takes place is a certain linear combination of
 $P_3$ and $\eps_6$. To determine it, a linear transformation
 $(P_3,\eps_6)\to(\xi,\zeta)$
is performed, which diagonalizes the quadratic form $(P_3,\eps_6)$
in the expansion (\ref{landau1}). Among
 $\xi$ and $\zeta$, the relevant order parameter is that one, which prefactor in
 (\ref{landau1}) vanishes at the Curie point.

 However, in practical calculations for the only experimentally accessible
field conjugate to the order parameter -- the electric field
 $E_3$, one usually uses the expansion in powers of polarization
 $P_3$
\begin{equation}
 \label{landau2}
 G=G_0+\frac{ a'(T-T_0)}{2}P^2+\frac{ b}{4}P^4+
\frac{ c}{6}P^6- EP, \end{equation}
and the higher order terms can be also taken into account.
For the sake of simplicity, in this section we omit indices near
polarization $P_3$ and field $E_3$.

In the vicinity of the transition temperature at $ E=0$, the
thermodynamic potential $G(P)$ has three minima $P_{1,3}^2=(-
b\pm\sqrt{ b^2-4 a c})/2 c$ and $P_2=0$, the depths of which become
equal at the transition point.  Hence, the temperature of the first  order
phase transition $T_{\rm C0}$ at $ E=0$ is found from the conditions
\[
\frac{\partial G}{\partial P}=0,\quad G(P_{1,3})=G(0),
\]
that are equivalent to
\[
3 b^2=16 a c, \quad
T_{\rm C0}=T_0+\frac{3 b^2}{16 c a'}.
\]

The extremum points of the thermodynamic potential $G(P)$ at $ E\neq0$
are given by the equation
\begin{equation}
\label{field}
 E=  aP+ b P^3 + cP^5.
\end{equation}
Among five solutions of this equations, two correspond to maxima,
and three $P_1<0$, $P_2\simeq0$, and $P_3>0$ correspond to minima
of $G(P)$. From the condition
\[
G(P_2)=G(P_3)~~{\rm at}~E>0, \quad  G(P_1)=G(P_1)~~{\rm at}~E<0,
\]
in a linear in $E$ approximation we find a field   dependence of
the first order phase transition temperature (the phase equilibrium curve)
\[
T_{\rm C}= T_{\rm C0}
- E\frac{12 c}{7a' b}\sqrt{\frac{4c}{-3b}}.
\]

\subsection{Critical point. Experimental measurements}
The phase equilibrium curves (the first order phase transition lines)
terminate in two critical points, which coordinates ($\pm E^*,T^*$) follow
from the thermodynamic conditions -- equation of state and the equilibrium
condition
 \[
  \left(\frac{\partial G}{\partial P}\right)_{ E}=0 ,\quad
  \frac{\partial  E}{\partial P}=0, \quad
  \frac{\partial^2  E}{\partial P^2}=0.
  \]
In these points a difference between
the truly ferroelectric and pseudo -``paraelectric'' with a non-zero
polarization $P_3$ and strain $\e$ phases disappears, so that
the transition from one to the other takes place smoothly without
jumps of the thermodynamic quantities.

Experimentally the coordinates of the critical points are usually found
using the phenomenological theory, by measuring in some way the coefficients
of the Landau expansion.  To determine the critical field
 $E^*$ and temperature $T^*$, the following
formulas are used  \cite{Schmidt},
\begin{eqnarray*}
&& E^*=a^*P^*+b(P^*)^3+c(P^*)^5+d(P^*)^7;\\
&& T^*=T_0+ \frac{a^*}{a'}=T_0-\frac{3b(P^*)^2+5c(P^*)^4+7d(P^*)^6}{a'},
\end{eqnarray*}
where the higher order terms in the expansion are also taken into account, and
$P^*$ depends on the number of terms in the expansion
\cite{Schmidt}
\begin{eqnarray*}
&& P^*=\frac{5c}{2d}\left[\sqrt{1-\frac{63}{25}\frac{bd}{c^2}}-1\right],
\qquad c\neq0, d\neq 0;\\
&& P^*=\sqrt{\frac{-b}{7d}}, \qquad c=0,d\neq0;\\
&& P^*=-\frac{3b}{10c}, \qquad c\neq0, d= 0.
\end{eqnarray*}

The simplest is to find the coefficient $a'$, which is nothing but
the inverse Curie constant of a crystal. The coefficients at higher
terms in the expansion are determined by  analyzing  the field and temperature
dependences of  the crystal polarization. Usually, the analysis of the so-called $\Gamma$-plots
is performed.
If one restricts the Landau expansion by the $P^6$ term, then from
the equation for field (\ref{field})
it follows \cite{Sidnenko2}, that the dependence of the quantity
\[
\Gamma(P^2) =\frac{E-aP}{P^3}=b+cP^2
\]
on  $P^2$ must be linear and independent of the
magnitude of the applied field. The intersection of this line with
the ordinate axis $(P^2)$ gives a value of $b$, and its slope gives
the coefficient $c$. Hence, by analyzing the experimentally measured
dependence $\Gamma(P^2)$ one is able to determine two coefficients of the
Landau expansion.

When the dependence $\Gamma(P^2)$ is closer to a quadratic one rather
than to to linear one, we should neglect the term $cP^5$ in the
Landau expansion and take into account the term $dP^7$, instead.
Then the dependence\[
\Gamma(P^4) =\frac{E-aP}{P^3}=b+dP^4
\]
must be linear.

Another, somewhat more sophisticated method of
determination of the Landau coefficients was proposed by Schmidt
\cite{Schmidt}. Instead of the $\Gamma$-plots, the so-called
isopols -- lines of a constant polarization in the $(E,T)$ plane should
be analyzed.  From (\ref{field}) it follows that
\[
T= \frac{E}{a'P}+T_0-\frac{bP^2+cP^4}{a'},
\]
that is, the dependence $T(E)$ at constant polarization is linear. Approximating
the experimental paraelectric isopols, one can find the coefficient $a'$
from the expression
\[
a'=\frac{1}{P}\left(\frac{\partial E}{\partial T}\right)_P.
\]
It should be noted that at the first order phase transition
the isopols do intersect, whereas at the second order phase
transition they do not. Therefore, this analysis
permits one to determine the order of the phase transition and
estimate the coordinates of the critical point (the point where
the order of the phase transition changes), if one exists.

Unfortunately, the described method of
determination of the critical point coordinates according to the
Landau coefficients  does not allow to find more or less
trustworthy values of the
critical quantities. In Table \ref{table-Schmidt} we presented the
measured in different papers Landau coefficients and calculated
 coordinates of the critical points for \kdp\ (the Table is taken from
\cite{Schmidt}). All the present results agree as for the
magnitude of the critical field in this crystal about
100-300~V/cm. We did not present here the data obtained within dilatometric
X-rays studies \cite{Kobayashi}, giving the critical field
in \kdp\ of about 8500V/cm.

\setlength{\tabcolsep}{0.48mm}

\begin{table*}[hbt]

\caption{
Landau coefficients and calculated coordinates of the critical point for
\kdp\ (the data are taken from \protect\cite{Schmidt}). }
\small
\begin{center}
\begin{tabular}{c|c|c|c|c|c|c|c}
\hline\noalign{\smallskip}
 {Ref}.
& $ a$
& $ b$
& $ c$
& $ d$
& $E^*$
& $T^*-T_0$
& method\\
&\footnotesize  ($ 10^{-3}$esu)
&\footnotesize  ($ 10^{-11}$esu)
&\footnotesize  ($ 10^{-19}\!$esu)
&\footnotesize  ($ 10^{-27}$esu)
& (V/cm)
& (K)
&\\
\noalign{\smallskip}\hline\noalign{\smallskip}
\cite{l366} 
& 3.9
& $-1.9$
& 6.3
& 0
& 120
& 0.07
&\\
 \cite{sugie} 
& $4.2\pm 0.1$
& $-1.9\pm 0.1$
& $5.4\pm 0.4$
& 0
& 160
& 0.07
&$\Gamma$-plot\\
 \cite{Schmidt19}
& $4.3\pm 0.2$
& $-2.35\pm 0.4$
&\footnotesize $5.91\pm 1.5$
& 0
& \footnotesize $232\pm 70$
& $0.10\pm 0.03$
& \footnotesize Isopols\\
 \cite{Schmidt20}
& $4.0\pm 0.2$
& $-1.48\pm 0.2$
& $3.1\pm 0.4$
& 0
& \footnotesize $186\pm 60$
& $0.08\pm 0.03$
& \footnotesize Isopols\\
 \cite{Schmidt}
&\footnotesize $3.91\pm 0.04$
&\footnotesize $-1.26\pm 0.05$
& $3.2\pm 0.1$
& 0
&\footnotesize $123\pm 18$
&\footnotesize $0.057\pm 0.007$
&\footnotesize Isopols\\
 \cite{Sidnenko3} 
& $3.8\pm 0.1$
& $-3.0\pm 0.8$
& $6.5\pm 1.1$
& 0
& $370$
& $0.16$
&$\Gamma$-plot\\
& $3.8\pm 0.1$
& $-0.5\pm 0.3$
& 0
&  $3.8\pm 0.4$
& $87$
& $0.036$
&$\Gamma$-plot\\
 \cite{Vallade} 
& $3.9$
&\footnotesize $-0.54\pm 0.05$
& 0
&\footnotesize $2.85\pm 0.10$
& $124$
& $0.046$
&\\
& $3.9$
&\footnotesize $-1.85\pm 0.25$
&  $3.3\pm 0.5$
&  $0.87\pm 0.5$
& $280$
& $0.11$
&\\
\noalign{\smallskip}\hline
\end{tabular}
\end{center}
\label{table-Schmidt}
\end{table*}
\normalsize

As we see, a considerable discrepancy between values of the Landau
coefficients takes place, and values of the critical point coordinates
are very sensitive to the choice of the Landau coefficients. Thus, a
1.5 times increase of the  $b$ coefficient leads to a
3 times as large value of the critical field
and twice as large the quantity $T^*-T_0$.
The  main sources of differences in the values of critical coordinates
determined via the Landau coefficients are the following
\begin{enumerate}
\item
Inadequacy of the Landau expansion as such. It is known to be
 suitable for the second order phase transitions, only.
For crystals, where the first order phase transition is observed,
a large number of terms in the expansion should be taken into account,
and even then it is valid only in the narrow vicinity of the
transition point. Even for \kdp, where the jump of polarization at the transition
point is not as large as in deuterated \dekdp, the Landau expansion is not
quite adequate.

\item
Ambiguity of the Landau expansion. At interpretation of the same
experimental data by different expansions \cite{Sidnenko3} (with $c=0$,
$d\neq0$ or with $c\neq0$,  $d=0$) the critical field can change to
more than 4 times. The expansion with $c=0$, $d\neq0$ yields considerably
smaller values of the critical field than the expansion with $c\neq0$,  $d=0$
does.

\item
A large numerical error in calculations of the expansion coefficients
at a given form of the expansion. Thus, at determining the coefficient
$b$ from $\Gamma$-plots, the error arises due to
the data dispersion for $\Gamma(P^2)$ at small values of polarization.

\item
Changes in the Landau coefficients from sample to sample. First,
a quite large dispersion of the  data (as for determination of the
critical point coordinates) takes place for the inverse Curie
constant (the coefficient $a'$) \cite{Sidnenko3}. Similarly,
the values of $b$ for different samples of the crystal  do differ,
even though they are obtained by the same method. Here, the dispersion
of the data may exceed the error of measurement
\cite{Schmidt,Schmidt19,Schmidt20}. Apparently,
the Landau coefficients and, thereby, the coordinates of the critical
point, essentially depend on a quality of the sample, its history,
presence of inner defects, etc.
\item
Different experimental methodics of polarization measurements
 (pyroelectric, dilatometric,
hysteresis loops) give values of the coefficient $b$, which may significantly
differ.
\end{enumerate}
Comparison with experimental data
of other quantities, calculated with the Landau expansion
 (for instance, temperature and field curves of polarization) does
not allow to ascertain these coefficients, because these other
quantities are not that sensitive to the values of the Landau
coefficients, as the coordinates of the critical point.

Hence, due to a large error in determination of the Landau coefficients
and ambiguity of the Landau expansion, the method of
calculation of the critical fields and temperatures
via the phenomenologic theory allows one only to {\it estimate}
the coordinates of the critical point (order of magnitude of the
critical field).

\section{Microscopic model}
\label{section1}

We consider a  ferroelectric crystal of the \kdp\ type to which external
electric field $E_3$, inducing  polarization $P_3$ and
strain $\e$ in the high-temperature phase, is applied.

The entire Hamiltonian of the model \cite{our!!} consists of a ``seed'' part,
independent of a hydrogen subsystem configuration and attributed to a host
lattice of heavy ions, and of pseudospin  short-range and long-range
 hydrogen Hamiltonians, tunneling being neglected
\be
H=N\bar v\left(
\frac{c_{66}^{E0}}{2}\e^2-e_{36}^{0}E_{3}\e-
\frac{\chi_{33}^0}{2}E_3^2\right)
+ H_{\rm long}+ H_{\rm short}.
\ee
The ``seed'' energy expressed in terms of the electric field $E_3$ and
strain $\e$ includes the elastic, piezoelectric, and electric
counterparts. $c_{66}^{E0}$, $e_{36}^{0}$, and  $\chi_{33}^0$ are the
so-called ``seed'' elastic constant, coefficient of the piezoelectric
stress, and dielectric susceptibility, respectively; $\bar v=v/{k_{\rm B}}$,
$v$ is the primitive cell volume; ${k_{\rm B}}$ is the Boltzmann constant;
$N$ is the number of primitive cells.

$H_{\rm long}$ is the  mean field Hamiltonian of the long-range
dipole-dipole and lattice mediated \cite{Blinc1966} interactions between
deuterons plus a linear in strain $\e$ molecular field \cite{stas1,our!!}
induced by piezoelectric coupling
\bea
&&
H_{\rm long}=\frac 12\sum_{q'f'qf}J_{ff'}(qq')
\frac{\langle\sigma_{qf}\rangle}{2}\frac{\langle\sigma_{q'f'}\rangle}{2}
- \sum_{qf}\left[
\left(\sum_{q'f'}J_{ff'}(qq')
\frac{\langle\sigma_{q'f'}\rangle}{2}\right)
\frac{\sigma_{qf}}{2} - 2\psi_6\e \frac{\sigma_{qf}}{2}\right]
 \nonumber\\ &&{}
= 2N\nu\eta^2 - \sum_{qf}\left(2\nu\eta-2\psi_6\varepsilon_6\right)
\frac{\sigma_{qf}}{2},
\eea
where
\[
4\nu={J_{11}+2J_{12}+J_{13}}
\]
is the eigenvalue of  the long-range interaction matrix Fourier transform
$J_{ff'}=\sum_{{\bf R}_q-{\bf R}_{q'}}J_{ff'}(qq')$;
\[
\eta=\langle\sigma_{q1}\rangle=\langle\sigma_{q2}\rangle=
\langle\sigma_{q3}\rangle=\langle\sigma_{q4}\rangle
\]
is the mean value of the Ising pseudospin $\sigma_{qf}=\pm 1$ which two
eigenvalues are assigned to two equilibrium
positions of a hydrogen on the $f$-th bond in the $q$-th unit cell.

The Hamiltonian  of  the short-range configurational interactions between
hydrogens is usually  chosen such as to reproduce the energy levels
of the Slater-type model for KDP (see, for instance Ref.\
\cite{Blinc_Zeks}) -- the energy levels  of up-down $\eps_s$ (twice
degenerate at $\e=0$ and $E_3=0$), lateral $\eps_a$  (four-fold
degenerate), single-ionized $\eps_1$ (eight-fold degenerate), and
double-ionized $\eps_0$ (twice degenerate)
hydrogen configurations
($\eps_s<\eps_a\ll\eps_1\ll\eps_0$).

Since the system is no longer symmetric with respect to the reflection
 $\sigma_h$  in the  $ab$ plane and mirror rotation $S_4$ around
the $c$ axis (both operations change the sings of polarization and
strain), in presence of strain $\e$ and in the electric field $E_3$ a
splitting of the energies of up and down ($i=1$ and $i=2$),
lateral ($i=5,6$) and ($i=7,8$), and single ionized ($i=9,10,11,12$ and
$i=13,14,15,16$) configurations takes place (confuguration numbers $i$
are given in Appendix).  Since strain $\e$ and polarization
$P_3$ transform after the same irreducible representation ($B_2$ in
the paraelectric phase and $A_1$ in the ferroelectric phase),
the field $E_3$ does not split those levels which remain degenerate
in presence of the strain $\e$.

That part of the splitting, which origin is the distortion of the
PO$_4$ groups and changes of the angle between perpendicular in the
paraelectric phase hydrogen bonds is described by linear
functions of the strain $\e$.  The contribution of the electric field
$E_3$ to the splitting is taken to be equal to the difference between
the energy levels of the groups dipoles in the
longitudinal field $E_3$.  Since the groups   K-D$_n$PO$_4$ with
hydrogens in up ($i=1$) and down ($i=2$) configurations and in
single-ionized configurations (configurations $i=9,10,11,12$ and
$i=13,14,15,16$) have oppositely directed projections of the dipole
moments $\boldsymbol{\mu}_i$ on the axis $c$, these groups have also
different energies $\boldsymbol{\mu}_i{\bf E}$ in external electric
field . Usually, the absolute values of
 the $c$-projections of the dipole moments of up/down configurations
$\mu_3$ are assumed to be twice as large as the corresponding
projections of single-ionized configurations. The $c$-projections of
dipole moments of lateral and double-ionized groups are zero.

Finally, the Hamiltonian of the short-range
interactions rewritten in terms of pseudospins according to standard
rules \cite{Blinc_Zeks,our!!} and taken into account within the  most
appropriate for these crystals four-particle cluster approximation, is
\bea
&&
 \hspace{-3.1ex}
\hat H_{\rm short}  = -
   \left[\Delta + \mu_3 E_3- \frac {2\delta_{16}-\delta_{s6}}4\e \right]
        \sum_q\sum_{f=1}^4\frac{\sigma_{qf}}2 + \sum_q \left\{
    U\left [
    \frac{\sigma_{q1}}2\frac{\sigma_{q3}}2+
    \frac{\sigma_{q2}}2\frac{\sigma_{q4}}2\right]+
    \Phi\frac{\sigma_{q1}}2\frac{\sigma_{q2}}2
    \frac{\sigma_{q3}}2\frac{\sigma_{q4}}2 +{}
  \right.  {} \nonumber\\
 &&  \hspace{-4.2ex}
  \label{2.4}
 -
\e (\delta_{s6}{+}2\delta_{16}) \!\!\left[\!
        \frac{\sigma_{q1}}2\frac{\sigma_{q2}}2\frac{\sigma_{q3}}2+
        \frac{\sigma_{q1}}2\frac{\sigma_{q2}}2\frac{\sigma_{q4}}2
  +
        \frac{\sigma_{q1}}2\frac{\sigma_{q3}}2\frac{\sigma_{q4}}2+
        \frac{\sigma_{q2}}2\frac{\sigma_{q3}}2\frac{\sigma_{q4}}2
\!      \right]  \!  \\
&&
\hspace{-3.1ex}
+\left.  (V+\delta_{a6}\varepsilon_6) \!\left [
		\frac {\sigma_{q1}}2\frac {\sigma _{q2}}2+
		\frac {\sigma_{q3}}2\frac {\sigma _{q4}}2\right]\!
 +
(V-\delta_{a6}\varepsilon_6)\!\left[
\frac{\sigma_{q2}}2\frac{\sigma_{q3}}2+\frac{\sigma_{q4}}2\frac{\sigma_{q1}}2
        \right]\right\} \nonumber.
\eea
Detailed derivation of the Hamiltonian and the table of the split energy levels
in presence of external electric field (longitudinal and transverse) are
given in Appendix.

In Eq.\ (\ref{2.4}) the following notations are used
\[
V=-\frac12w_1, \quad
 U=\frac12w_1-\varepsilon,\quad
\Phi=4\varepsilon+2w_1-8w;
\]
where
 \[
\varepsilon=\varepsilon_a-\varepsilon_s,\quad
w=\varepsilon_1-\varepsilon_s,\ \ w_1=\varepsilon_0-\varepsilon_s
 \]
are the so-called Slater energies. Splitting of the energy levels due
to lowering the system symmetry is the only changes in the
short-range interactions that we take into account within this model.
Therefore, $\eps$, $w$, $w_1$ do not depend on strain
$\e$ of field $E_3$.

The condition of equality of the mean values of pseudospins
$\eta=\langle\sigma_{qf}\rangle$  calculated with the four-particle
(\ref{2.4}) and single-particle deuteron Hamiltonians (see \cite{our!!})
permits to exclude the self-consistency parameter $\Delta$.
The order parameter $\eta$ and strain $\e$ can
 be found by minimization of the thermodynamic potential (Gibbs' function)
 $g_{1E}(T,\s,E_3,\eta)$
\begin{eqnarray}
&& g_{1E}(T,\s,E_3,\eta)=
\frac{\bar v}{2}c^{E0}_{66}\varepsilon^2_6 -
{\bar v}e^0_{36}\varepsilon_6E_3
- \frac{\bar v}{2}\chi^{\varepsilon 0}_{33}E^2_3
+ 2T\ln 2 + 2\nu\eta^2 -
2T\ln(1- \eta^2)D- {\bar v}\sigma_6\varepsilon_6,
\nonumber
\end{eqnarray}
where
\begin{eqnarray*}
&&
D = \cosh (2z +\beta\delta_{s6}\e) +
     4b \cosh (z -\beta\delta_{16}\e)+       aa_6
     + \frac{a}{a_6}+d,\nonumber\\
&&z = \frac12\ln\frac{1+\eta}{1-\eta} +
\beta\nu
\eta - \beta\psi_6\varepsilon_6 + \frac{\beta\mu_3E_3}{2},
\\
&& a = \exp(-\beta\varepsilon), \; b = \exp(- \beta w), \;
d= \exp(-\beta w_1), \;
 a_6 = \exp(- \beta\delta_{a6}\varepsilon_6).
\nonumber
\end{eqnarray*}
 The conjugate to the strain
$\e$ stress $\s$ is introduced for the sake of simplicity; in
numerical calculations $\s=0$.

Using the system of equations
\begin{eqnarray*}
&&
\sigma_6 = c^{E0}_{66}\varepsilon_6 -
e^0_{36}E_3 + \frac{4\psi_6}{\bar v}
\frac{m }{D}
+\frac{2M_6}{\bar v D },
\\
&&
P_3 = e^0_{36}\varepsilon_6 + \chi^{\varepsilon 0}_{33}E_3 +
2\frac{\mu_3}{v}\frac{m }{D},
\end{eqnarray*}
which follows from the thermodynamic equilibrium conditions
\[
\frac{1}{\bar v}\left( \frac{\partial g_{1E}}{\partial\varepsilon_6}
\right)_{T,E_{3},\sigma_6} = 0, \qquad
\frac{1}{\bar v}\left( \frac{\partial g_{1E}}{\partial E_3}
\right)_{T,\sigma_6} = - P_3
\]
we find
the isothermal  dielectric susceptibility of a free crystal
(at $\sigma_6 ={}$const)
\be \label{first} \chi^{T\sigma}_{33}
= \left(\frac{\partial P_3}{\partial E_3} \right)_{T\sigma_6} =
\chi^{T\varepsilon}_{33} + \frac{(e_{36}^T)^2}{c_{66}^{TE}},
\ee
expressed via the isothermal
 dielectric susceptibility of a clamped
crystal (at $\e={}$const)
\be \chi^{T\varepsilon}_{33} =
\left(\frac{\partial P_3}{\partial E_3}\right)_{T\varepsilon_6} =
\chi^0_{33} + \frac{\mu_3^2}{v} \frac{2\beta\varkappa}{D-
2\varkappa\varphi},
\ee
the isothermal coefficient of piezoelectric
stress
\be e_{36}^T =   \left( \frac{\partial
P_3}{\partial\varepsilon_6}\right)_{T,E_3} = e^0_{36} + \frac{2\mu_3}{v}
\frac{\beta\theta }{D-2\varkappa\varphi},
\ee
and the isothermal elastic constant at constant field
\bea
&&c^{TE}_{66}=\left(
\frac{\partial\sigma_6}{\partial\varepsilon_6}\right)_{T,E_3} = c^{E0}_{66}
+ \frac{8\beta\psi_6}{{\bar v}}
  \frac{-\varkappa\psi_6+r }{D -
2\varkappa\varphi} +
\frac{2\beta}{{\bar v}D^2} M_6^2
 \nonumber\\  &&  {}
- \frac{2\beta}{{\bar v}D} \left[
\delta_{s6}^2\cosh (2z +\beta\delta_{s6}\e)+
\delta_{a6}^2(aa_6+\frac{a}{a_6})+
4b \delta_{16}^2\cosh (z -\beta\delta_{16}\e)\right]
-
\frac{4\varphi r^2}{{\bar v}TD(D -
2\varkappa\varphi)}.\nonumber
\eea
The other isothermal dielectric and piezoelectric characteristics
can be recalculated via the found above by using the known formulas

\noindent  the constant of piezoelectric stress $h_{36}$
   \be
   h_{36} = - \left( \frac{\partial
E_3}{\partial\varepsilon_6}\right)_{TP_3} = - \left(
\frac{\partial\sigma_6}{\partial P_3}\right)_{T\varepsilon_6} =
\frac{e^T_{36}}{\chi^{T\varepsilon}_{33}}
\ee
the
elastic constant $c^{P}_{66}$ at  constant polarization
\be
c^{P}_{66} = \left(
\frac{\partial\sigma_6}{\partial\varepsilon_6}\right)_{TP_3} = c^{TE}_{66} +
e_{36}^Th_{36}.
\ee
\noindent compliance at constant field
\be
s^{TE}_{66} =
\left(\frac{\partial\varepsilon_6}{\partial\sigma_6}\right)_{TE_3}
= \frac{1}{c^{TE}_{66}},
\ee
the coefficient of the piezoelectric strain
\be
d_{36}^T=\left(\frac{\partial P_3}{\partial\sigma_6}\right)_{TE_3} =
e^T_{36}s_{66}^{TE},
\ee
the constant of piezoelectric strain
\be
\label{last}
g_{36} = - \left(\frac{\partial E_3}{\partial\sigma_6}\right)_{TP_3} =
 \frac{h_{36}}{c^{TP}_{66}},
\ee

When dynamic experimental methodics  are used (measuring frequency exceeds
the thermal relaxation frequency), not the isothermal but adiabatic
characteristics of the crystals are measured. In the paraelectric phase
in absence of electric field the adiabatic and isothermal quantities
coincide, but they may essentially differ in the ferroelectric phase or
in the electric field. The adiabatic quantities can be obtained
from the isothermal analogs with the help of the following
formulas \cite{Kanzig}
\begin{eqnarray} && \label{hiadia}
\frac{1}{\chi_{33}^{S\sigma}}=\frac{1}{\chi_{33}^{T\sigma}}\left[1+
\frac{1}{\chi_{33}^{T\sigma}}
\frac{(\partial P_3/\partial T)_{E_3}^2}{(\partial S/\partial T)_{P_3}}
\right]=
\frac{1}{\chi_{33}^{T\sigma}}\left[1+
\frac{T}{\chi_{33}^{T\sigma}}
\frac{p_\sigma^2}{c_{P_3,\sigma}}
\right];\nonumber\\
&&
d_{36}^{S}=d_{36}^{T}\frac{\chi_{33}^{S\sigma}}{\chi_{33}^{T\sigma}};\\
&&
(c^{SE}_{66})^{-1}=(c^{TE}_{66})^{-1}-
\left(1-\frac{\chi_{33}^{S\sigma}}{\chi_{33}^{T\sigma}}\right)
\frac{(d_{36}^{T})^2}{\chi_{33}^{T\sigma}},
\nonumber
\end{eqnarray}
where
\[
S =
R \bigl( - 2\ln2 + 2\ln [1 - \eta^2] + 2\ln D +
4T\varphi^T\eta + \frac{2\bar M_6}{D} \bigr)
\]
is the entropy of the hydrogen subsystem,
$p_\sigma$ is the pyroelectric coefficient,
$c_{P_3,\sigma}$ is the specific heat at constant polarization and
pressure.
 From
(\ref{hiadia}) it follows that the most prominent difference
between the adiabatic and isothermal quantities takes place in
the range of rapid temperature changes of polarization (where
$(\partial P_3/\partial T)_{E_3}^2$ is large) -- in the vicinity of
the transition point or of the permittivity maximum.

Using (\ref{hiadia})  and the relations
(\ref{first})-(\ref{last}) we can show that the adiabatic and
isothermal ``true''  constants of the crystals -- the elastic constant
at constant polarization $c_{66}^P$ and piezomodules
$h_{36}$ and $g_{36}$ -- coincide.

The pyroelectric coefficient and the contribution of the hydrogen
 subsystem into the specific heat at constant pressure and
polarization are given by the following expressions
 \be p_6^\sigma =
p_6^\varepsilon + e_{36}\alpha_6,
\ee
and
\be
\label{4.3}
\Delta
c_{P_3,\sigma} = \frac T{v}\Bigl( \frac{\partial S}{\partial
T}\Bigr)_{\sigma} =
\Delta c_{P_3\varepsilon} + \frac{q_6^P\alpha_6}{v}, \ee
where
$\Delta  c_{P_3\varepsilon}$ is the specific heat at constant
polarization and strain
\[ \hspace{-2ex} \Delta c_{P_3,\varepsilon} =
\frac Tv \left(\frac{\partial S}{\partial T} \right)_{P_3,\varepsilon_6} =
\frac{2R}{D}\left\{ 4T\varphi^T [\varkappa T\varphi^T
+ \tilde q_6 ] + N_6 - \frac{{{\bar M}_6}^2}{D} \right\},
\]
$R$ is the gas constant; $q_6^P$ is the strain heat at constant $P_3$,
\[ \hspace{-2ex}
q_6^P =
  \left(\frac{\partial S_6}{\partial\varepsilon_6}\right)_{P_3,T} =
\frac{4R}{D}\left\{ T\varphi_6^T
(r-2\varkappa_6\psi_6 ) -
\tilde q_6
\psi_6
\lambda_6+\frac{{{\bar M}_6}M_6}{2D}
\right\},
\]
$\alpha_6$ is the factor produced
by temperature variation of the strain $\e$
\[ \alpha_6 = \left(\frac{\partial \varepsilon_6}{\partial
T}\right)_{\sigma}= \frac{-q_6^P + h_{36}p_6^{\varepsilon}}{c_{66}^E}
\]
(true thermal expansion, which invokes the diagonal strains, is not
considered here. We used the following notations
\begin{eqnarray*} &&
\hspace{-5ex} N_6 =\frac{1}{T^2}\!\left[\!
({\varepsilon+\delta_{a6}\e})^2aa_6+
({\varepsilon-\delta_{a6}\e})^2\frac{a}{a_6}+
4 b{w}\delta_{16}\e \sinh(z -\beta\delta_{16}\e)
+\right.\nonumber\\
&&
\hspace{-5ex}
\left. +
4 b[w^2+\delta^2_{16}\e^2 ] \cosh(z -\beta\delta_{16}\e)
+ {w_1}^2 d-
\left({\delta_{s6}\e}\right)^2 \cosh(2z +\beta\delta_{a6}\e)\right];
\nonumber\\
&&
\hspace{-5ex}
\tilde q_6 = \frac{1}{T}\!\left[
-\delta_{s6}\e\cosh(2z +\beta\delta_{as}\e)+
  2 b\delta_{16}\e\cosh(z -\beta\delta_{16}\e)
+  2bw \sinh(z -\beta\delta_{16}\e) \right] -
 \eta \bar M_6,
\nonumber \\
&&
\hspace{-5ex}
\lambda_6=\frac{1}{T}\!\left[\!
-\delta_{s6}^2\e\sinh(2z +\beta\delta_{a6}\e)
+
 \delta_{16}^2\e 4 b\cosh(z_6 -\beta\delta_{16}\e)
\!+\right.\nonumber\\
&&
\hspace{-5ex}
\left. +
\delta_{a6}\! \left(\!aa_6(\varepsilon + \delta_{a6}\varepsilon_6) -
\frac{a}{a_6}
(\varepsilon - \delta_{a6}\varepsilon_6)\!\right)+
  \delta_{16}w\e 4 b\sinh(z -\beta\delta_{16}\e)\right].
  \nonumber\\
&&
\hspace{-5ex}
\varphi^T = - \beta^2(\nu \eta -
\psi_6\varepsilon_6), \\
&&
\hspace{-5ex}
{{\bar M}_6} = \beta (4b w \cosh z + d w_1 +
a\varepsilon a_6 +a\varepsilon a_6^{-1}+M_6\e).
\end{eqnarray*}
$R$ is the gas constant.

Having the pyroelectric coefficient and the specific heat at constant
polarization, we can find the total molar specific heat at constant pressure
\be
\label{specheat}
 \Delta
c_6^\sigma = T\left( \frac{\partial S}{\partial T}\right)_{\sigma} =
\Delta c_6^{P\sigma} + q_6^{\varepsilon} p_6^{\sigma}.
\ee
\[
 q_6^{\varepsilon} = \left(\frac{\partial S}{\partial P_3}
\right)_{\varepsilon_6,T} = \frac{v}{\mu_3}\frac{2RT}{D}\varphi
\{ 2\varkappa T\varphi^T + \tilde q_6\}
\]
is the polarization heat at given $\varepsilon_6$,

\section{Discussion}
\subsection{Fitting procedure}
The presented in previous Section theoretical results will
be used to description of the physical characteristics of a highly
deuterated crystal \dekdp. We shall also verify whether the presented
theory is capable of describing behavior of the physical
characteristics of an undeuterated
\kdp\ without taking into account tunneling effects.

For a highly deuterated crystal K(H$_{1-x}$D$_x)_2$PO$_4$ with
the transition temperature in zero field $T_{\rm C0}=211.73$~K
(a nominal deuteration level $x=0.89$, hereafter abbreviated as \dekdp) we
use the values of the theory parameters found in \cite{our!!}, providing
a fair quantitative description of temperature of several associated with
strain $\e$ dielectric, piezoelectric, and elastic characteristics
 of the crystal at atmospheric pressure, namely, the dielectric permittivities
of a free and clamped crystals, elastic constants  $c_{66}^E$ and
$c_{66}^P$ of open-circuited and short-circuited crystals, and piezomodules
$e_{36}$, $d_{36}$, $g_{36}$ and $h_{36}$. Details of the fitting procedure
are given in \cite{our!!}.

For and undeuterated crystal \kdp\ we also chose the theory
parameters such as to fit the temperature behavior of
the elastic constants $c_{66}^E$ and $c_{66}^P$, piezomodules $e_{36}$,
$d_{36}$, $g_{36}$, and $h_{36}$, as well as the longitudinal
static dielectric permittivity of a clamped crystal
$\eps_{33}^\eps$. It should be noted that
only the data of \cite{a12,a13} for the temperature dependences
of the coefficient of piezoelectric strain
$d_{36}$, of \cite{l147,l370,l379,l380} for the dielectric permittivity
$\eps_{33}^\sigma$, and of \cite{a14} for the
elastic constant $s_{66}^E$ are direct experimental points for \kdp.
Using Eqns.\ (\ref{first}) -- (\ref{last}) and having the values of
$d_{36}$, $\eps_{33}^\sigma$, and $c_{66}^E$, we  find
``experimental''  points for the piezoelectric constants $e_{36}$,
$h_{36}$, $g_{36}$, elastic characteristics $c_{66}^P$, $s_{66}^E$,
$s_{66}^P$, and clamped dielectric permittivity $\chi_{33}^\varepsilon$.
Recalculated thus values of the
piezoelectric constants $g_{36}\sim 53\cdot 10^{-8}$~cm$^2$/esu and
$h_{36}\sim 3.8\cdot 10^4$~dyn/esu above the transition point
are greater than $44\cdot 10^{-8}$~cm$^2$/esu and
$3.0\cdot 10^4$~dyn/esu, respectively, given by Mason \cite{Mason_old}
but closer to the deQuervain values ($50\cdot 10^{-8}$~cm$^2$/esu for
$h_{36}$).

By this set of the parameters we also try to
describe the field dependences of polarization $P_3$ \cite{Chabin},
longitudinal static dielectric susceptibility
$\chi_{33}^{S\sigma}$ and elastic constant $c_{66}^{SE}$ of a \kdp, obtained
in \cite{Litov}. It should be noted that in \cite{Litov}
only the elastic constant was measured directly by the ultrasonic technique,
whereas the susceptibility was recalculated using the known thermodynamic
relations and the data of \cite{deQuirvane} for the piezomodules.

A scheme of choosing the theory parameters for \kdp\ was analogous to that for
\dekdp.
At chosen $\eps$ and $w$, close to those used in theories where the piezoelectricity
and tunneling effects are not taken into account \cite{Zach1}, and setting
the Curie temperature of a clamped crystal $T_{\rm 0}^{\rm clamp}=118.65$~K,
we found the value of the long-range interaction parameter $\nu$.
Setting the values of the deformation potentials $\psi_6$, $\delta_{a6}$, $\delta_{16}$,
we found the value   $\delta_{s6}$ by fitting to experimental
value of the piezomodule $e_{36}$ at a certain temperature.
The parameters
 $\eps$, $ w$, $\psi_6$, $\delta_{a6}$, $\delta_{16}$, $c_{66}^{E0}$,
 $\mu_3/v$ were chosen such as to obtain the correct values of
the transition temperature  $T_{\rm C0}=122.8$~K \cite{Chabin}
of a free crystal,
ratio of the polarization jump at the transition point
to saturation polarization (jump of the order parameter), correct
temperature curves of the elastic constants $c_{66}^{E}$ \cite{a14} and
$c_{66}^{P}$, piezomodules
$d_{36}$, $g_{36}$, and $h_{36}$, longitudinal
static dielectric permittivities of a free $\eps_{33}^\sigma$
\cite{l379,l384,l147,l370,l375}  and
clamped $\eps_{33}^\eps$  crystals, as well as
specific heat. Properly chosen parameters $e_{36}^{0}$, $\chi_{33}^{(0)}$
yield the correct values of the piezomodules
$d_{36}$, $e_{36}$, $g_{36}$, and $h_{36}$ in the high-temperature limit.

In Table~\ref{parameters} we present the adopted values of
the theory parameters for \dekdp\ and \kdp.

\begin{table}[hbt]
\caption{The theory parameters.}
\small
\begin{center}
\begin{tabular}{|l|ccccccc|cccc|}
\hline
 &  $ \eps$
& $ w$
& $ \nu$
& $\psi_6$
& $\delta_{s6}$
& $\delta_{a6}$
& $\delta_{16}$
&  $c_{66}^{E0}\cdot 10^{-10}$
& $\displaystyle\frac{\mu_3}{v}$
& $e_{36}^{0}$
& $\chi_{33}^{(0)}$\\
& \multicolumn{7}{c|}{(K)}
&(dyn/cm$^2)$
& ($\mu$C/cm$^2$)  & (esu/cm$^2$) &
\\
\hline
\dekdp & 91.3 & 781 & 34.615
& $-500$ & $-692.5$ & 1350 & 50 & 6.7 &  6.9 & $0.42\cdot 10^4$ & 0.4
\\
\kdp\ & 52.5 & 383 & 20.58
&-380 &  $-257$ & 300  & 40 & 7.3 &  6.4 &
  $0.3\cdot 10^4$ &
  \\
\hline
\end{tabular}
\end{center}
\label{parameters}
\end{table}

\subsection{Effective dipole moment}
In our calculations, as well as in earlier theories (see, for
instance, \cite{Chabin}) two different values of the effective dipole
moment $\mu_3^-<\mu_3^+$ must be used in the ferroelectric and paraelectric
phases.
The value of $2\mu_3^-/v$ in the ferroelectric
phase is given by the experimental value of
saturation polarization.
The paraelectric values of $2\mu_3^+/v$, listed in Table~\ref{parameters},
are set by the relation
between the experimental and theoretical values of the Curie constant
for the longitudinal static dielectric permittivity. In the present theory,
$\mu_3^+$ must also provide an agreement with experiment for the
piezomodules
\[
e_{36}\sim {\mu_3},\quad
d_{36}\sim {\mu_3},\quad
h_{36}\sim {\mu_3}^{-1},\quad
g_{36}\sim {\mu_3}^{-1}
\]
in the paraelectric phase. In \kdp\ and \dekdp\ the ratio $\mu^+/\mu^-$ is
about 1.2. Neither in the present model nor in simpler versions of the
proton
ordering models it is possible to choose the theory parameters such as
to fit the temperature curves of polarization and permittivity  with
a single value of the parameter $\mu_3$.

The necessity to use in calculations two different values of $\mu$ for
paraelectric and ferroelectric phases is not a problem while we do not
consider behavior of the system in external fields conjugate to the
order parameter. These fields smear out the phase transition; therefore,
when the difference between the phases becomes only quantitative (in
fields above the critical one), it does not make sense to distinguish
the paraelectric and ferroelectric phase and the corresponding $\mu^+$ and
$\mu^-$.

Tokunaga \cite{Rhodes} relates the deviation of the ratio $\mu^+/\mu^-$
from unity with the existence of an underdamped soft mode  in a crystal.
In the displacive type ferroelectrics $\mu^+/\mu^-\gg1$, whereas in
the order-disorder type ferroelectrics $\mu^+/\mu^-\sim1$.
Failure of the order-disorder model to describe both polarization and
dielectric permittivity without invoking two values of $\mu$ for different
phases shows the limits of the model suitability. Since the \kdp\ family
crystals undergo  phase transitions of a  mixed ordering-displacive type
(it is believed that hydrogen ordering triggers displacements of heavy
ions), for a consistent description of dielectric properties of these
crystals, we need to sophisticate the model and include phonon degrees of
freedom and anharmonicity into consideration. Moreover, we need to go
further than
Kobayashi model (mixed proton-phonon model with only one optical mode
taken into account), which is known only to renormalize parameters of the
proton ordering model.

\subsection{Zero field case}
\begin{figure}[p]
\begin{center}
\leavevmode
\epsfysize=4.5cm
\rotate[r]{{\epsffile{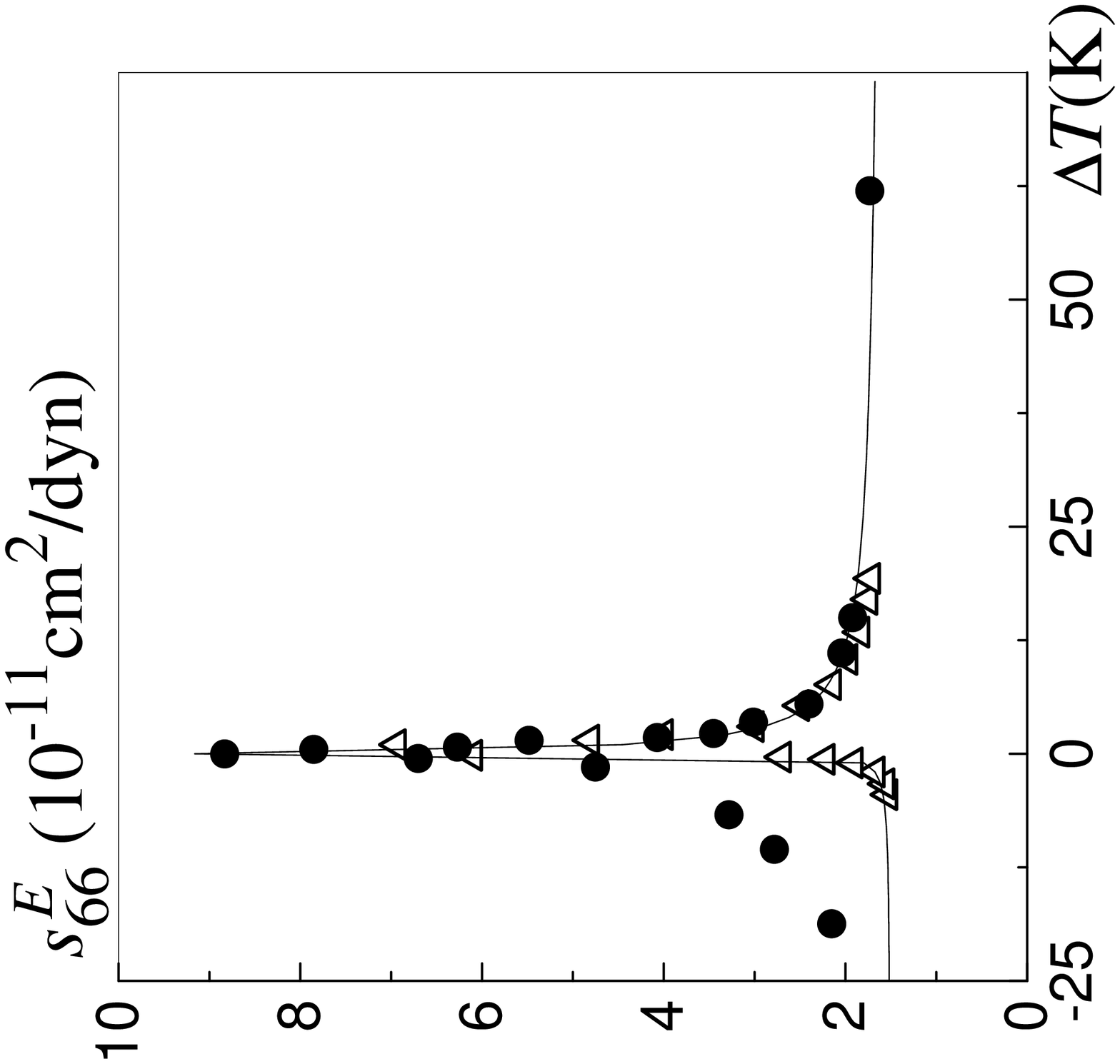}}}
\hspace{0.5cm}
\epsfysize=4.5cm
\rotate[r]{{\epsffile{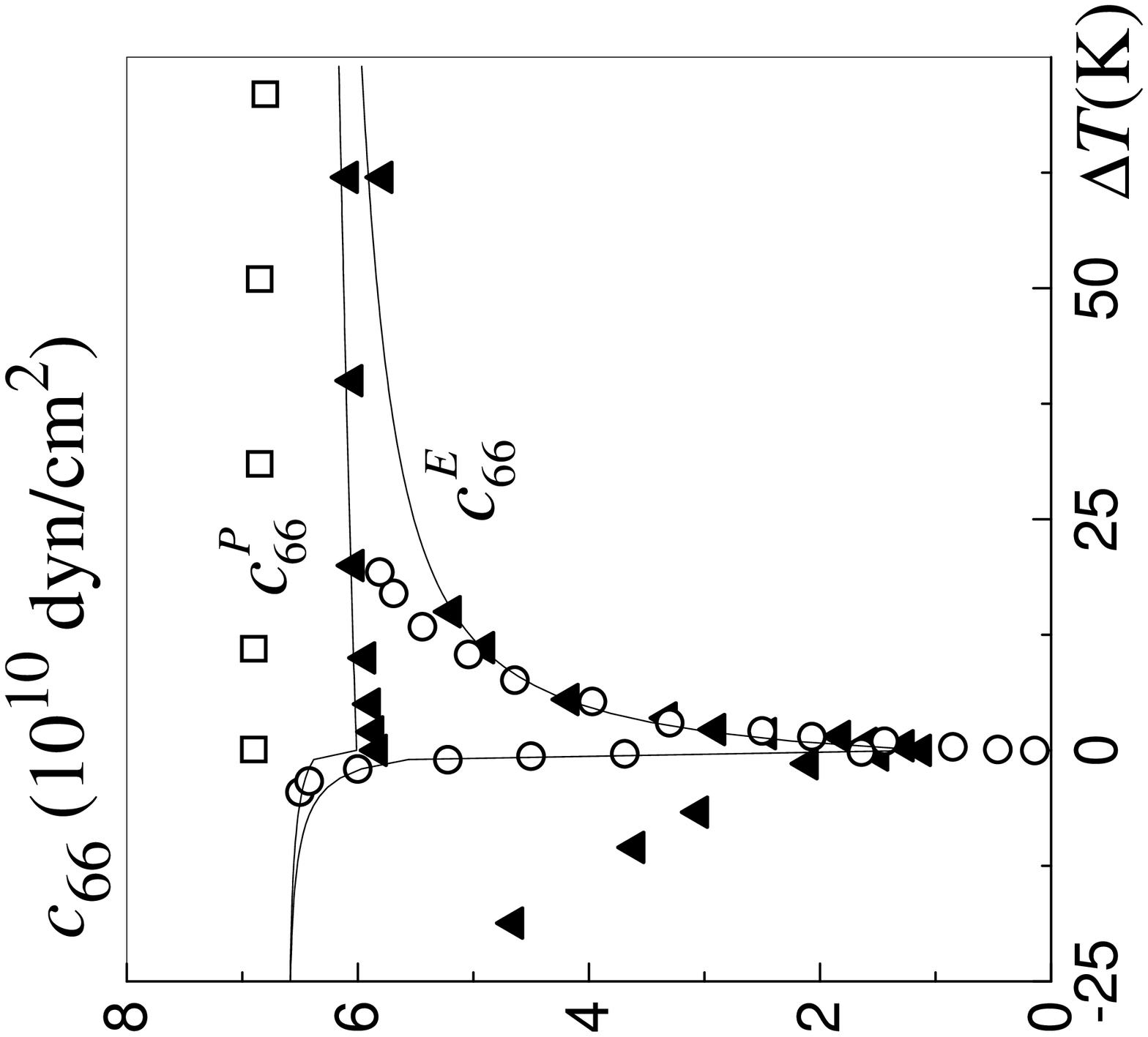}}}

\noindent
\epsfysize=4.5cm
\rotate[r]{{\epsffile{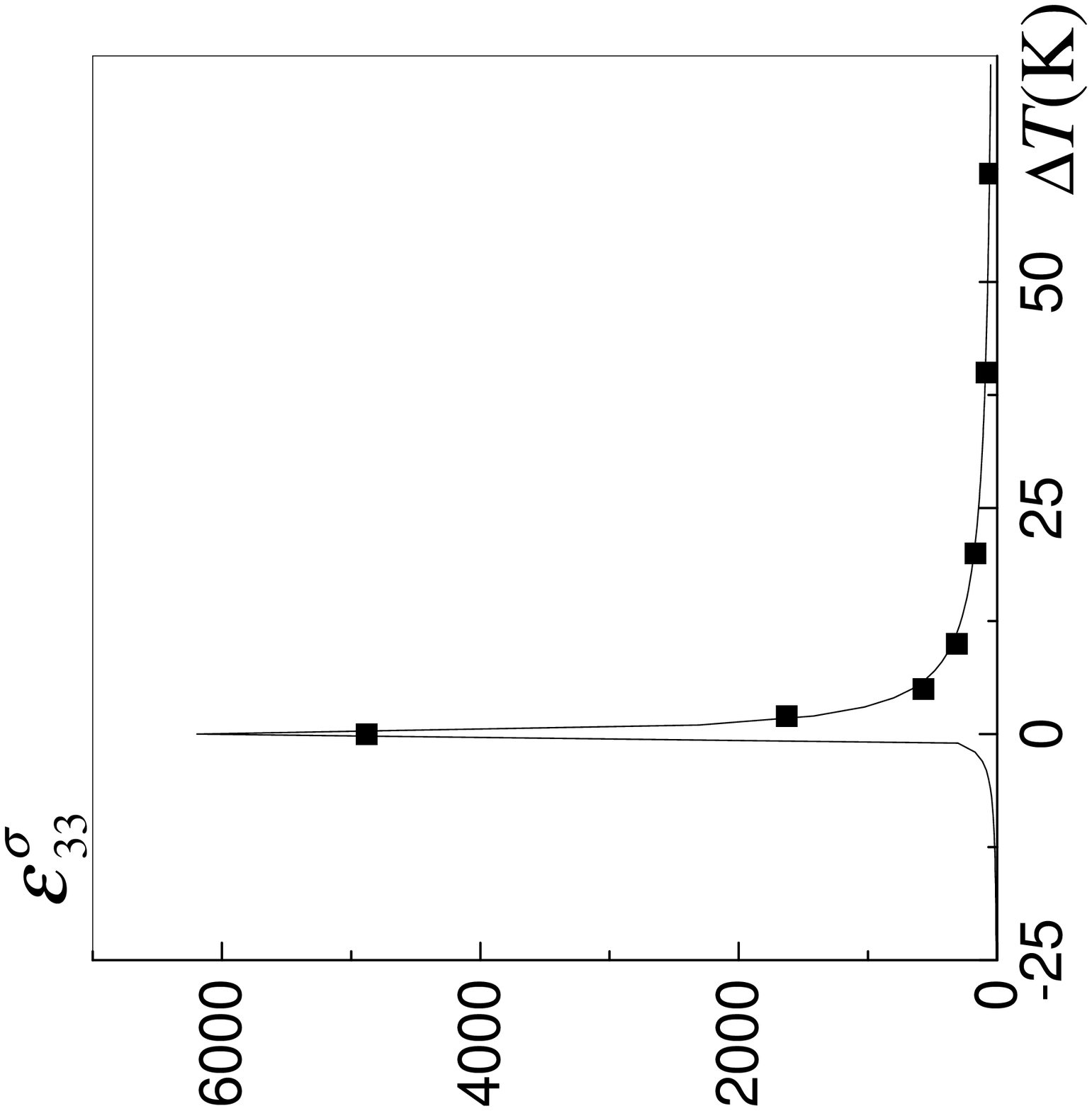}}}
\hspace{0.5cm}
\epsfysize=4.5cm
\rotate[r]{{\epsffile{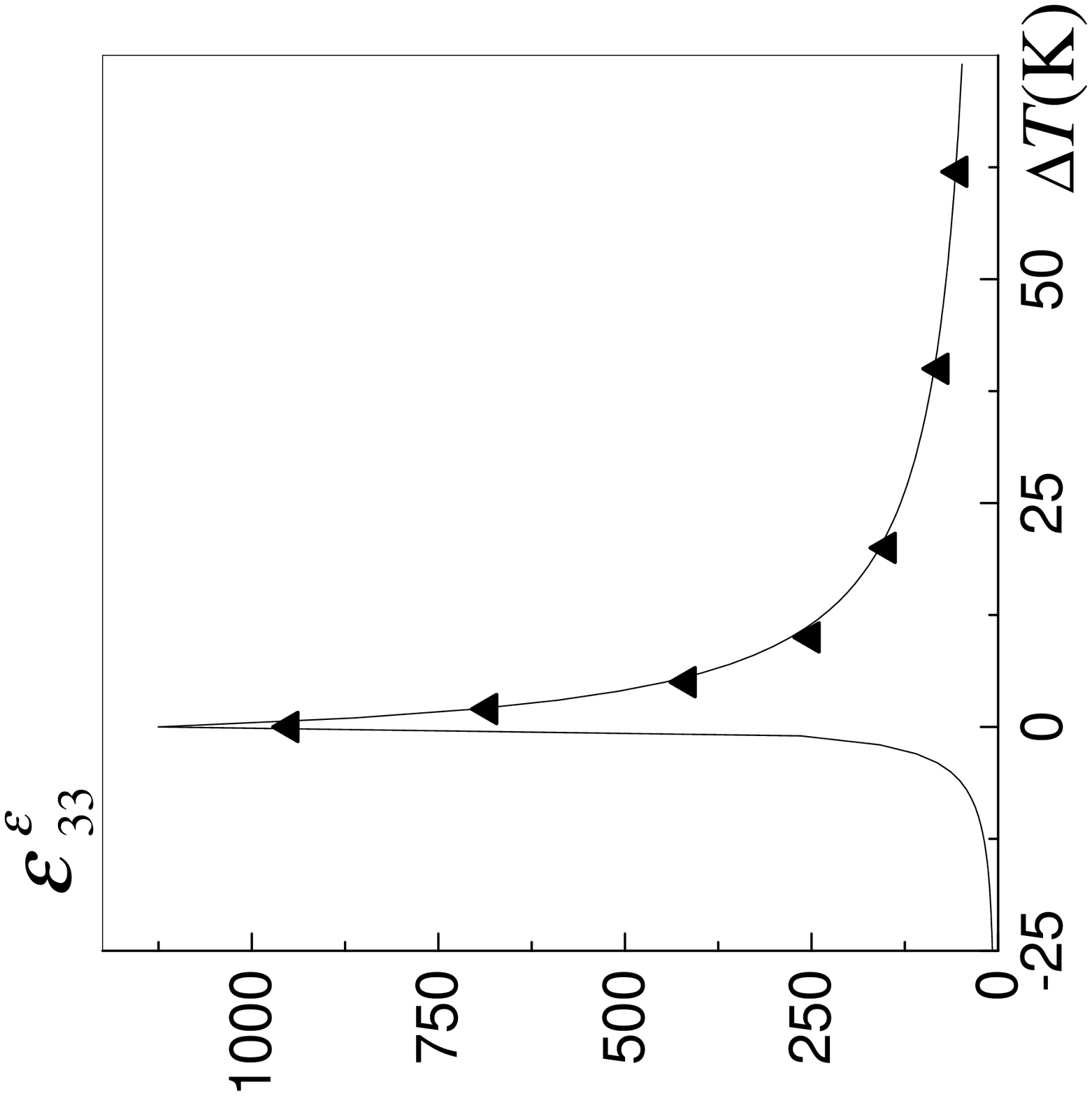}}}

\noindent
\epsfysize=4.5cm
\rotate[r]{{\epsffile{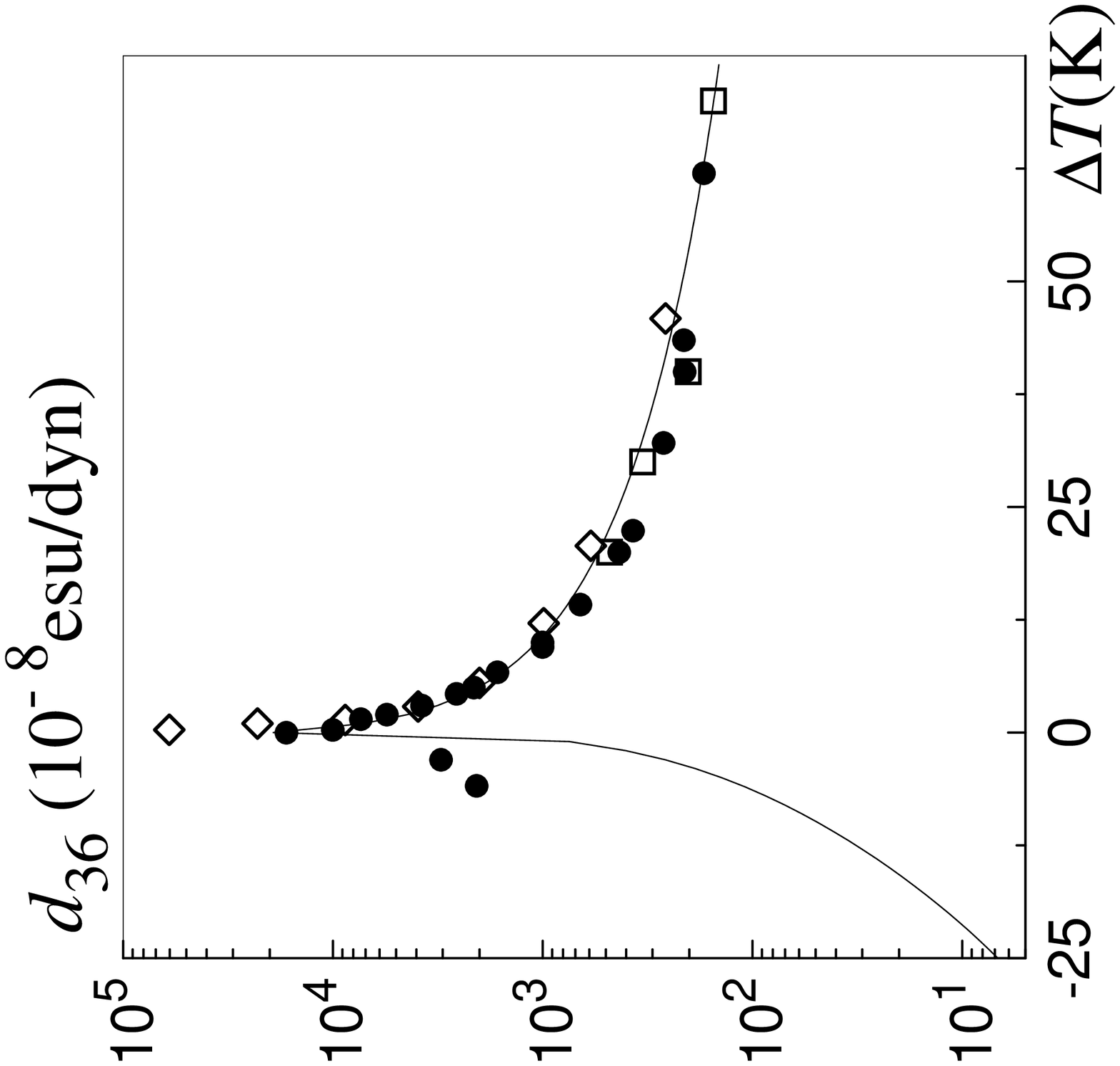}}}
\hspace{0.5cm}
\epsfysize=4.5cm
\rotate[r]{{\epsffile{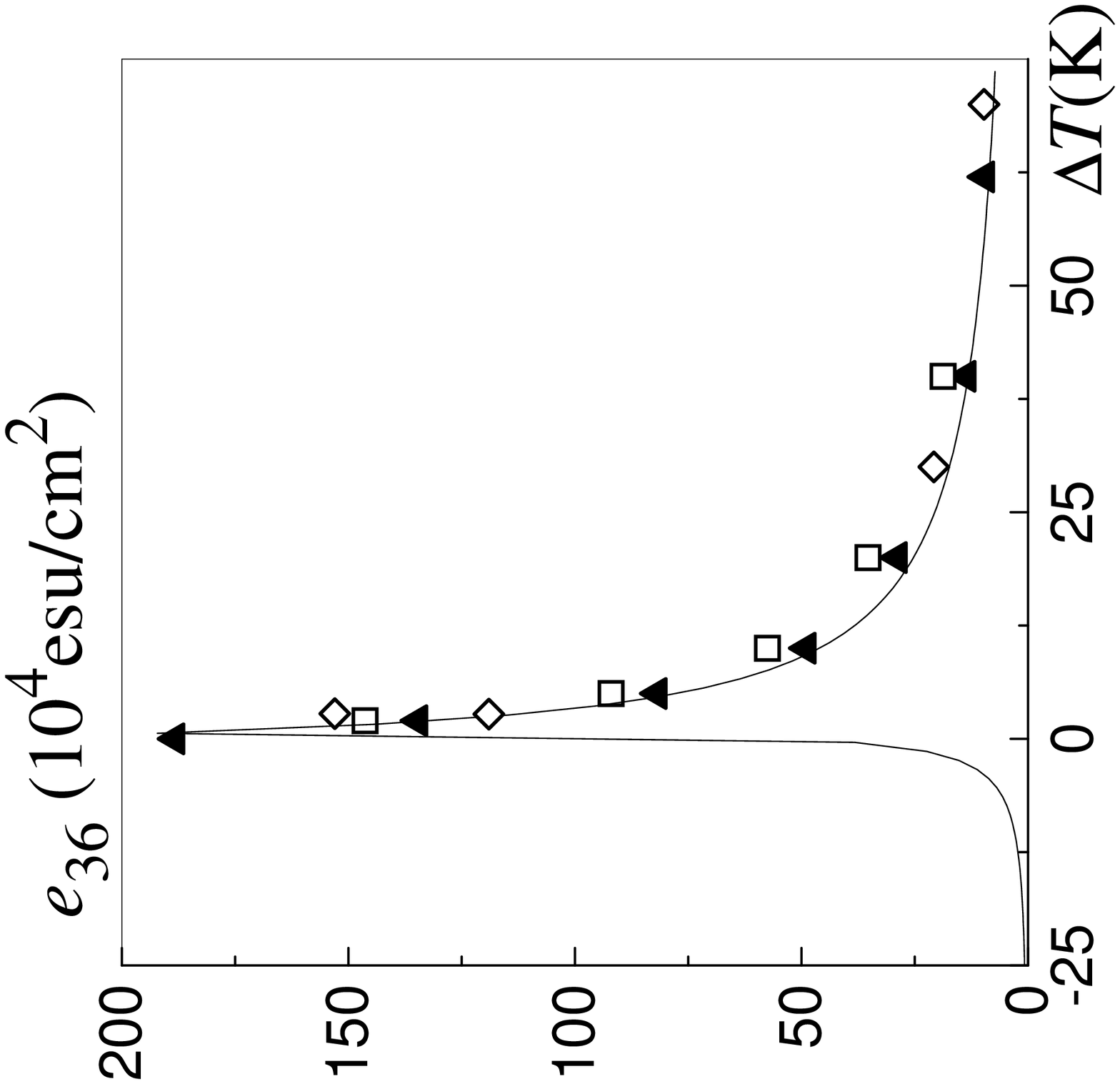}}}

\noindent
\epsfysize=4.5cm
\rotate[r]{{\epsffile{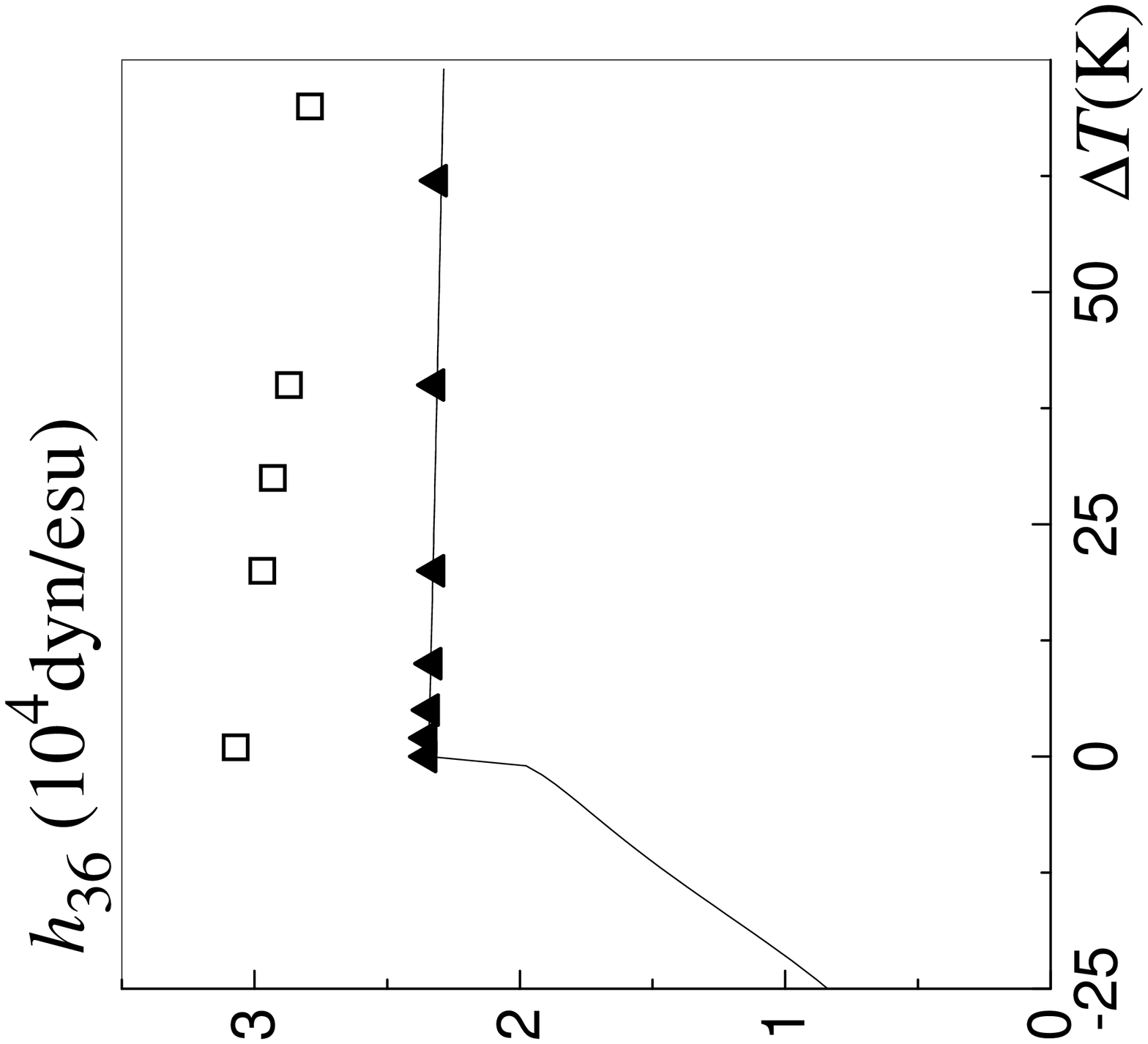}}}
\hspace{0.5cm}
\epsfysize=4.5cm
\rotate[r]{{\epsffile{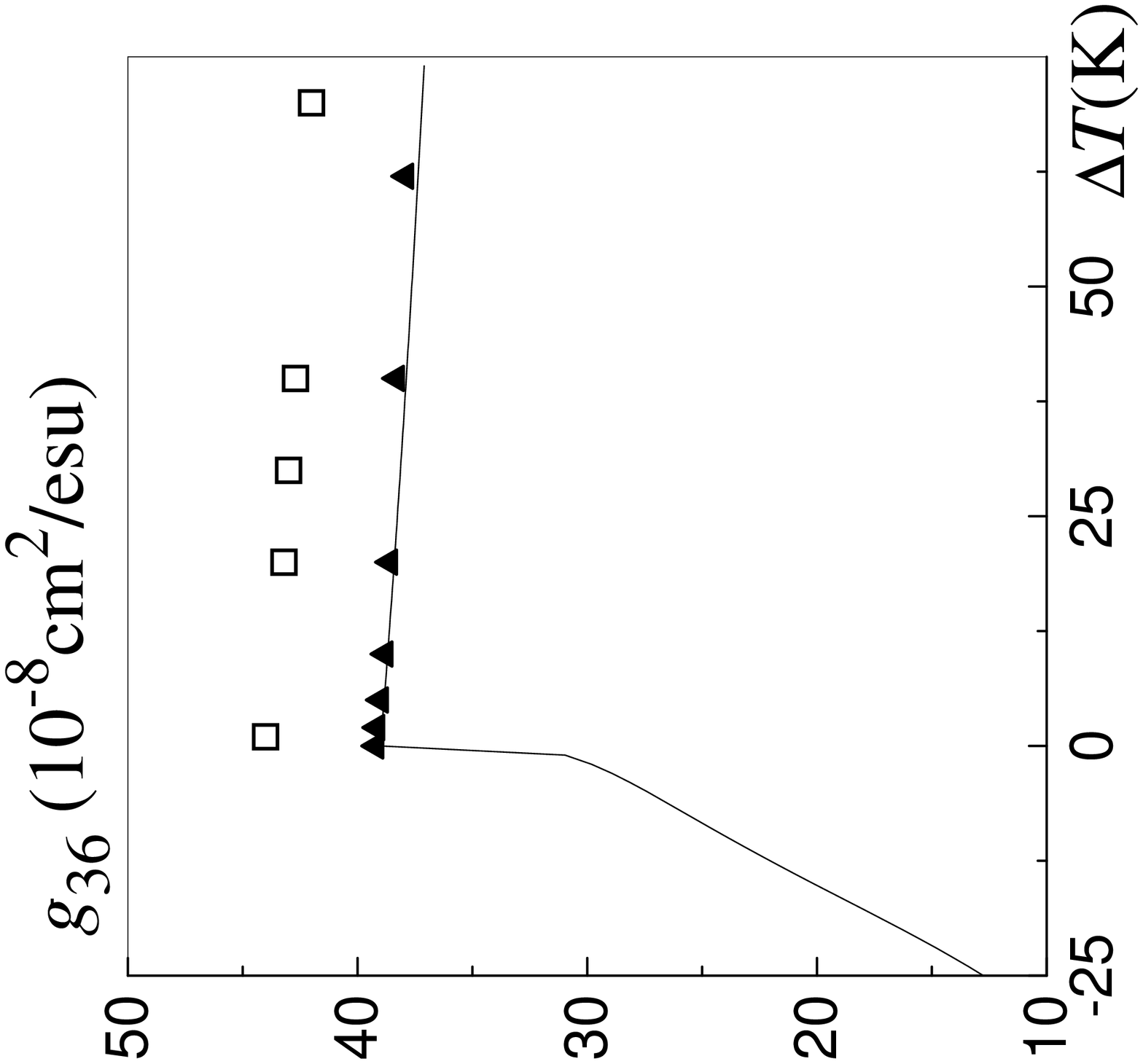}}}
\end{center}
\vspace*{-3ex}
\caption{Temperature dependences of the strain $\e$-related physical
characteristics. Lines and solid symbols correspond to a deuterated \dekdp;
open symbols representing pure \kdp\ are shown for comparison. Experimental
points are taken from
$\square$ -- \protect\cite{Mason_old};
$\circ$ -- \protect\cite{a14};
$\diamond$  -- \protect\cite{a12,a13};
$\blacksquare$ -- \protect\cite{a15};
$\bullet$ -- \protect\cite{Shuv};
$\triangle$ and $\blacktriangle$ are recalculated from Eqns.
(\ref{first}) -- (\ref{last})
 using experimental data of Refs.\ \protect\cite{Mason_old,a14} and
 Refs.\ \protect\cite{a15,Shuv}, respectively.}
  \label{zeroth} \end{figure}

\begin{figure}[p]
\begin{center}
\leavevmode
\epsfysize=4.5cm
{\rotate[r]{\epsffile{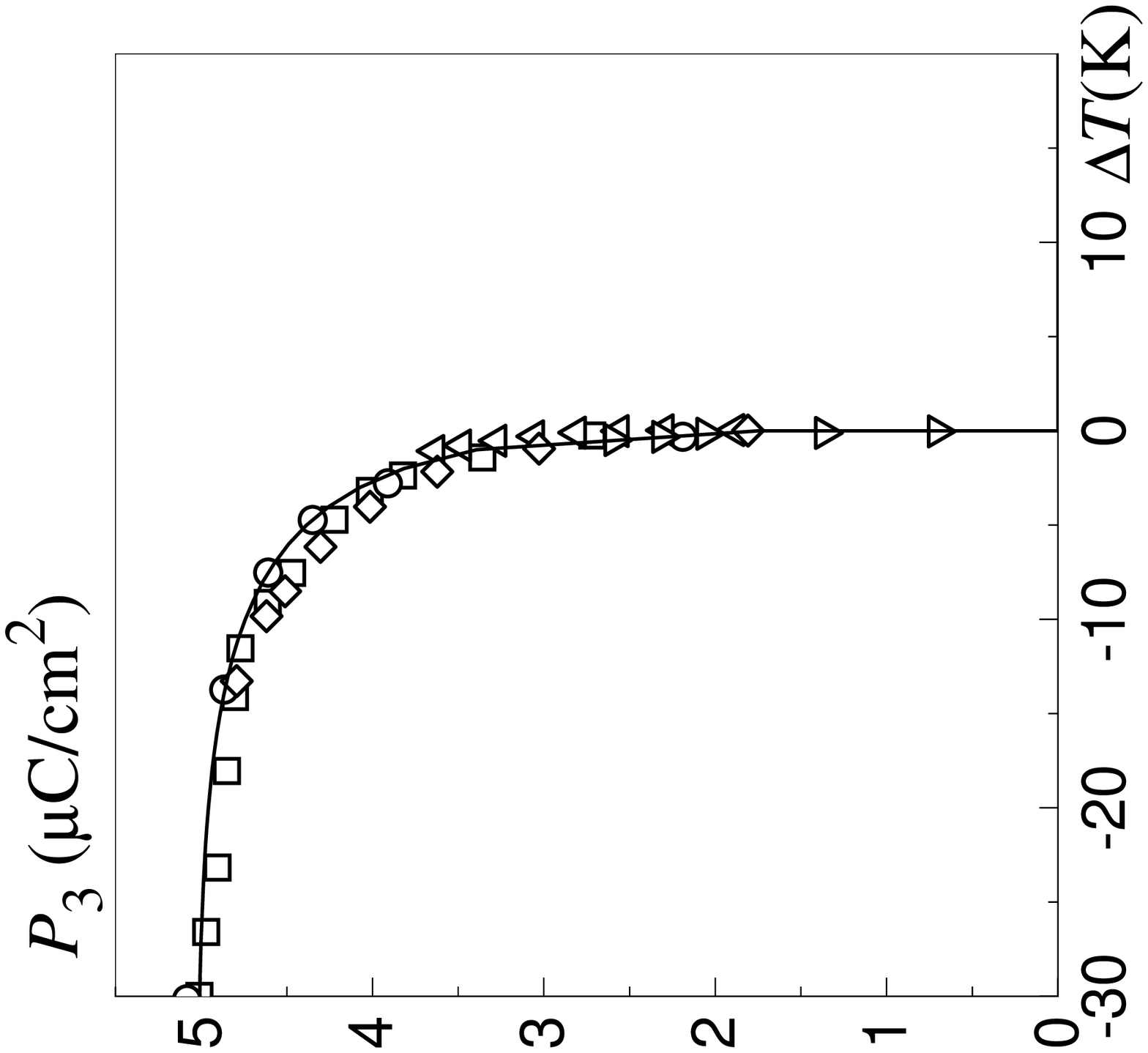}}}
\hspace{0.5cm}
\epsfysize=4.5cm
{\rotate[r]{\epsffile{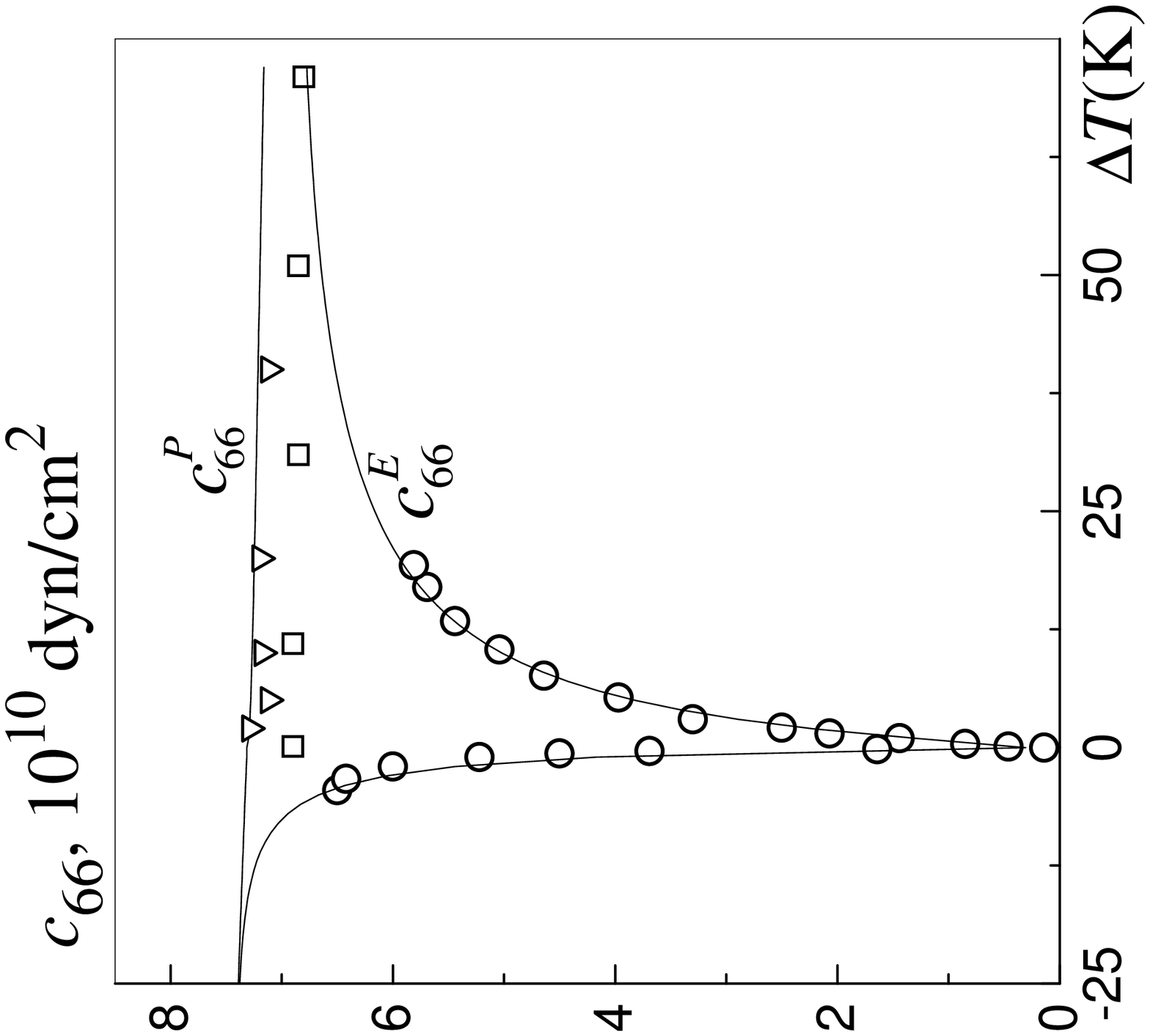}}}

\noindent
\epsfysize=4.5cm
{\rotate[r]{\epsffile{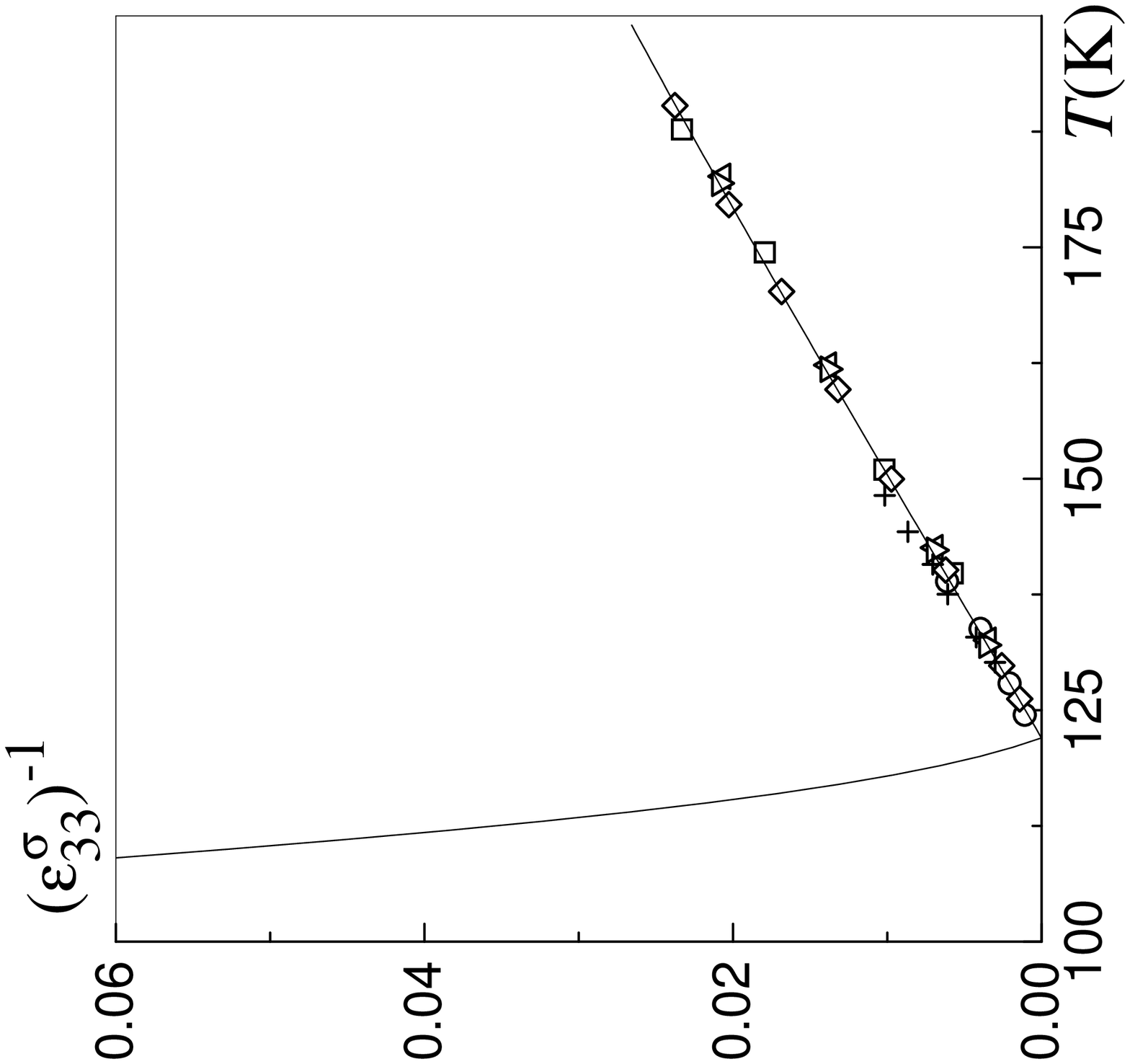}}}
\hspace{0.5cm}
\epsfysize=4.5cm
{\rotate[r]{\epsffile{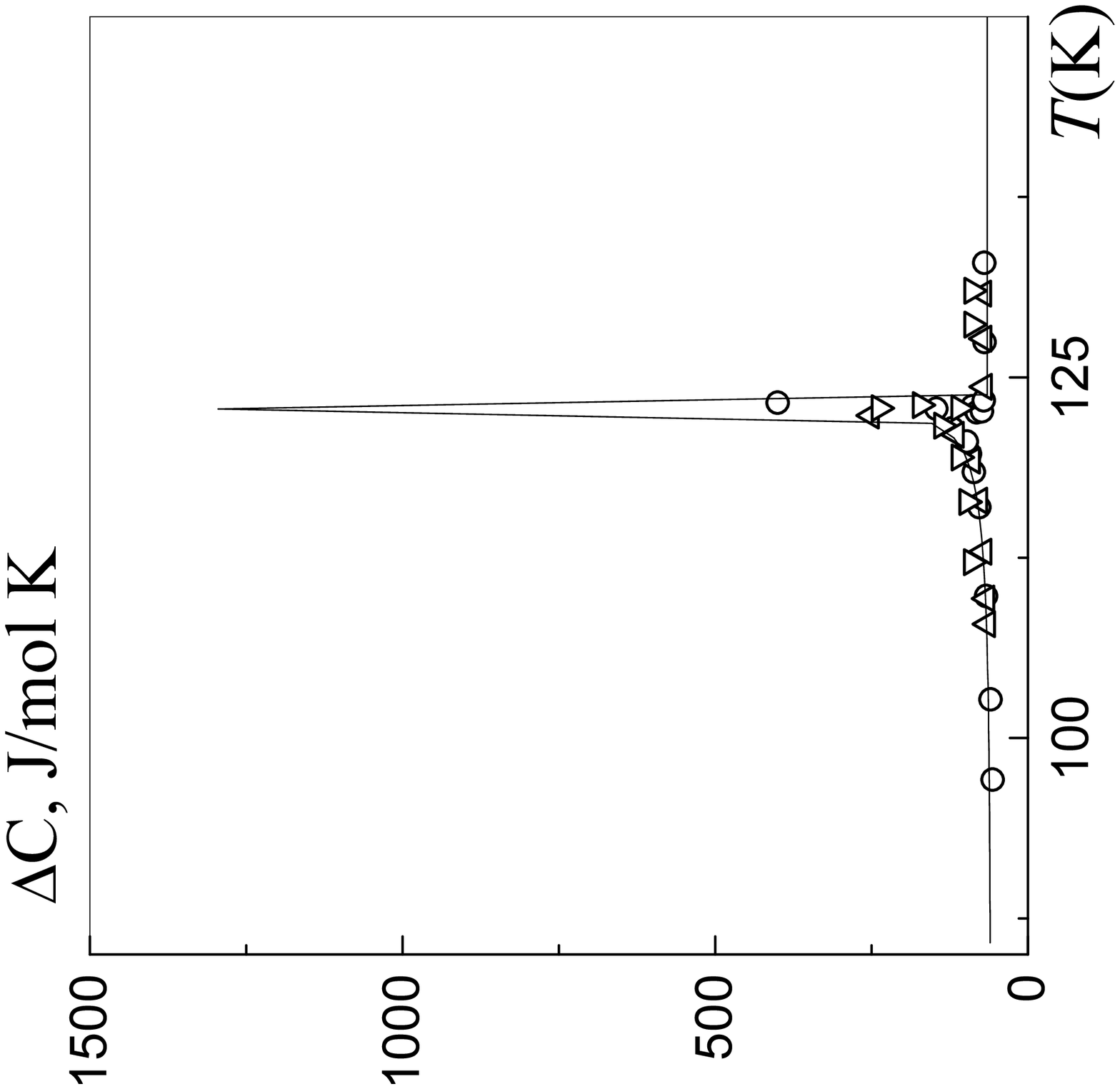}}}

\noindent
\epsfysize=4.5cm
{\rotate[r]{\epsffile{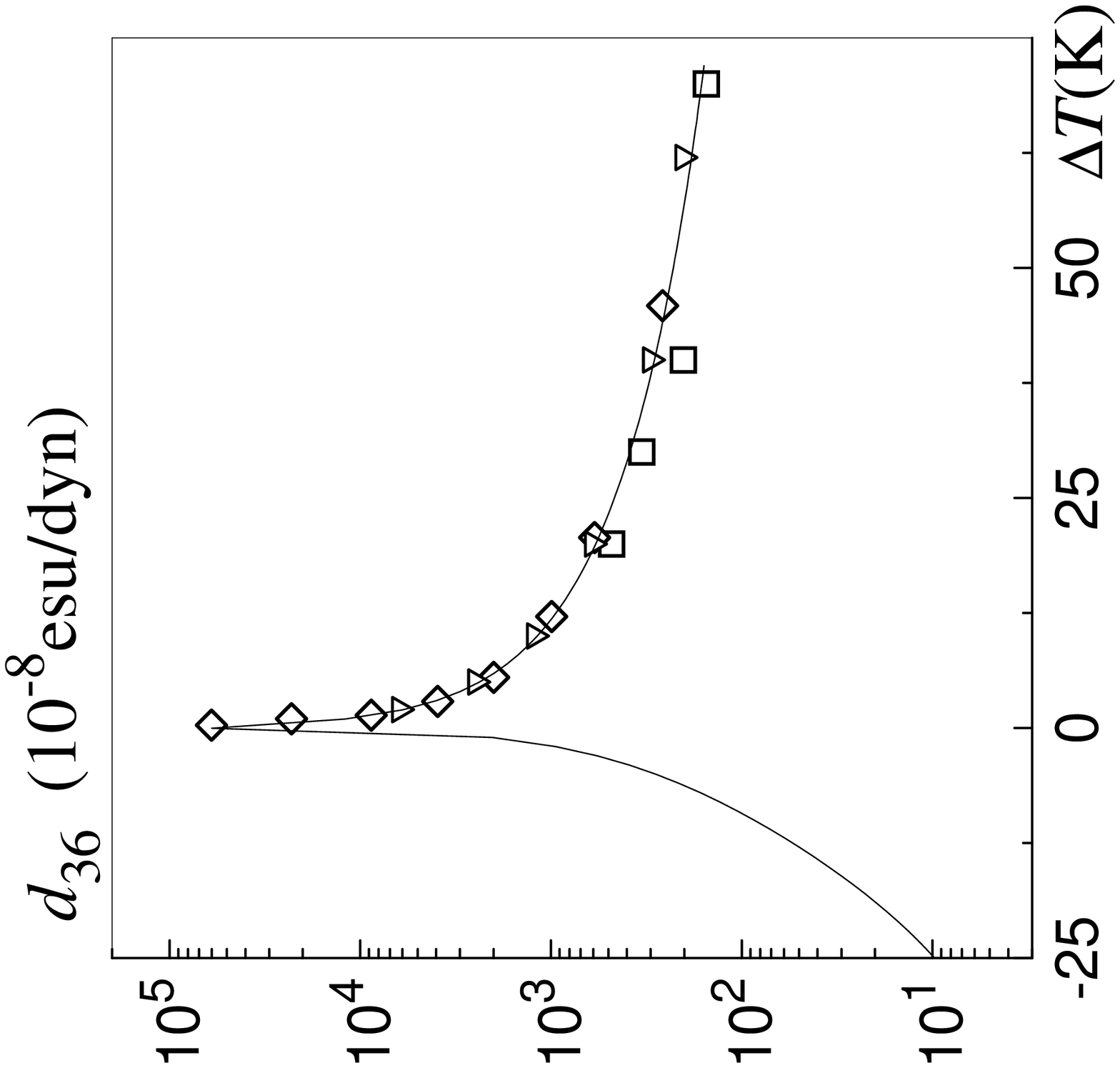}}}
\hspace{0.5cm}
\epsfysize=4.5cm
{\rotate[r]{\epsffile{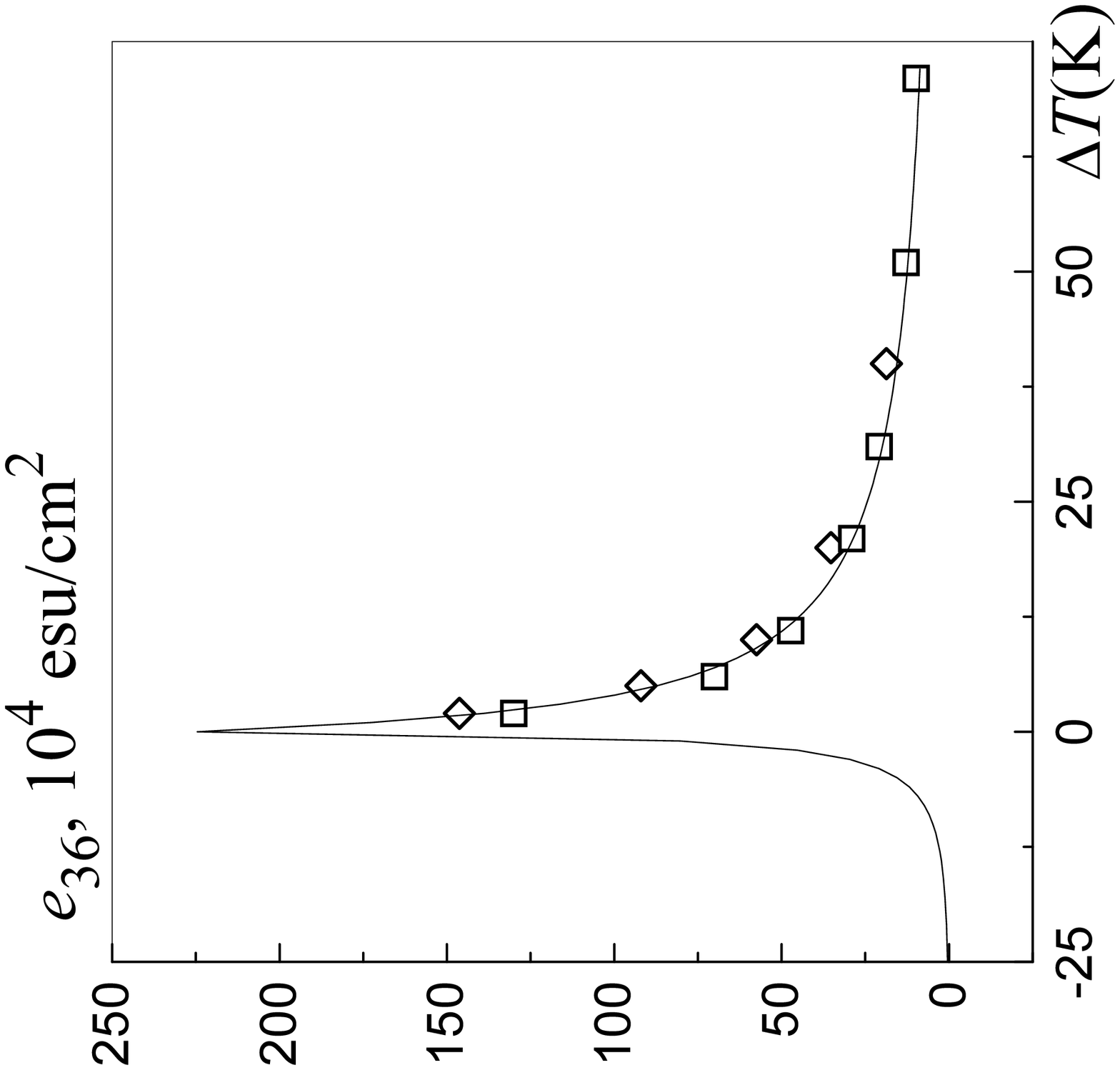}}}

\noindent
\epsfysize=4.5cm
{\rotate[r]{\epsffile{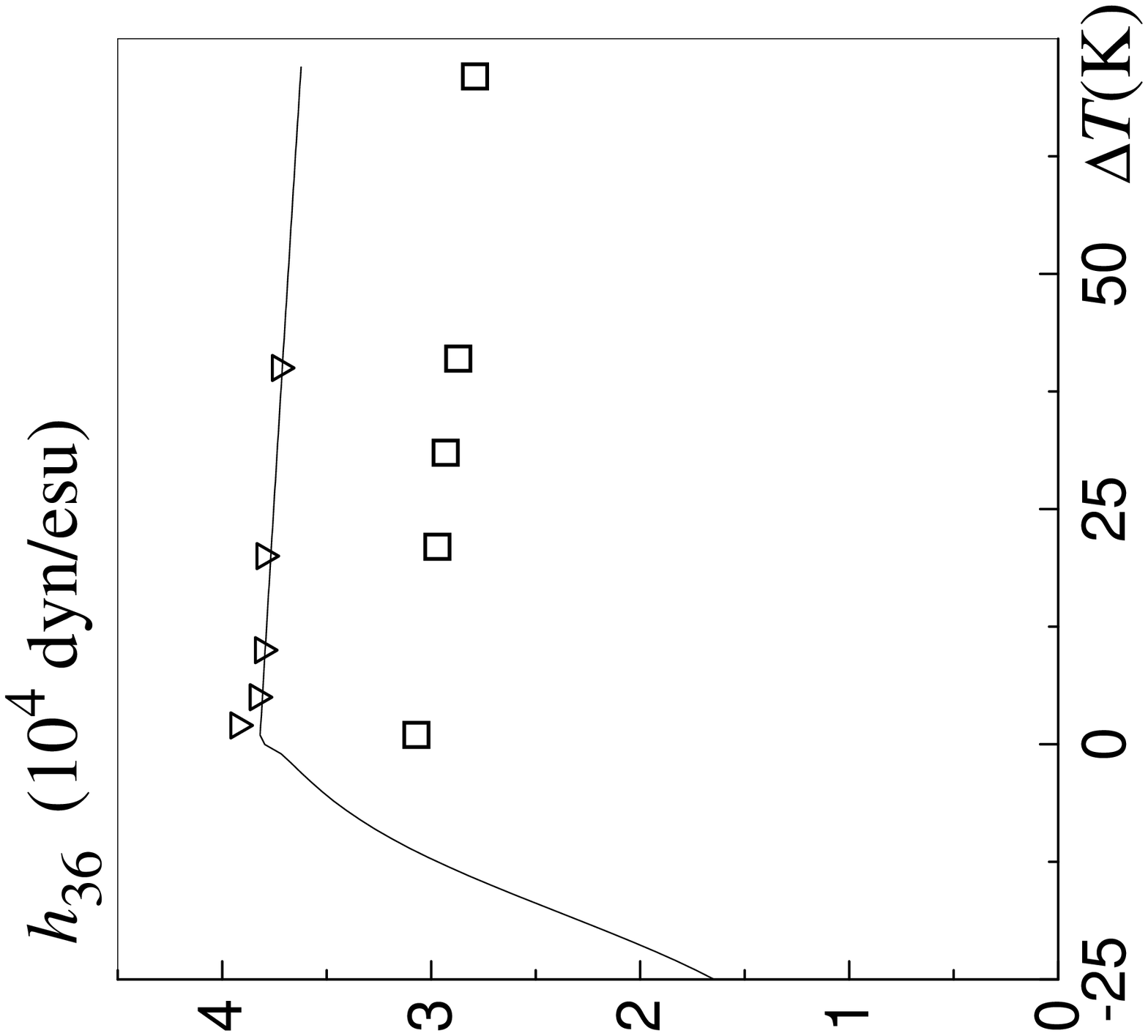}}}
\hspace{0.5cm}
\epsfysize=4.5cm
{\rotate[r]{\epsffile{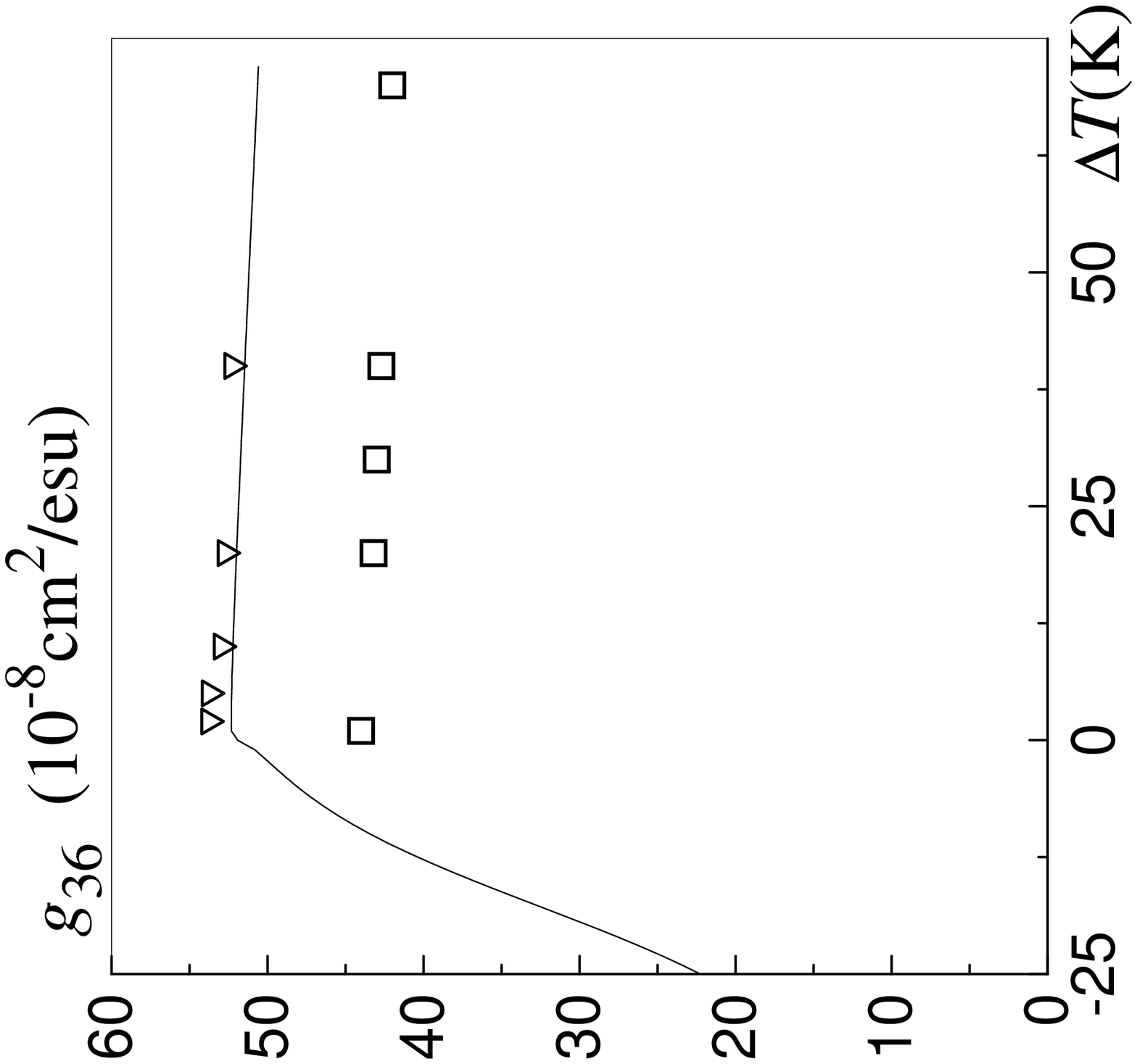}}}
\end{center}
\caption{Temperature dependences of the physical
characteristics of an undeuterated crystal \kdp. Experimental
points:
for polarization:
$\square$ -- \protect\cite{l147};
$\circ$ -- \protect\cite{l363};
$\triangle$ -- \protect\cite{l365};
$\triangledown$ -- \protect\cite{l366};
for the permittivity:
$\square$ -- \protect\cite{l379};
$\circ$ -- \protect\cite{l384};
$\triangle$ -- \protect\cite{l147};
$\triangledown$ -- \protect\cite{l370};
$\diamond$ -- \protect\cite{l375};
for specific heat:
$\square$ -- \protect\cite{l412};
$\circ$ -- \protect\cite{l417};
$\diamond$ -- \protect\cite{l380};
for the others:
$\square$ -- \protect\cite{Mason_old};
$\circ$ -- \protect\cite{a14};
$\triangledown$ are recalculated from Eqns.
(\ref{first}) -- (\ref{last})
 using experimental data of Refs.\ \protect\cite{a12,a13,a14,l379}.
}
  \label{zeroth2} \end{figure}

Figures \ref{zeroth} and \ref{zeroth2} illustrate how the proposed theory
describes experimental data for the physical characteristics of
\dekdp\ and \kdp\ crystals in absence of external electric field.
Polarization of a pure \kdp\ $P_3$
is calculated with  $2\mu_3^-/v=5.0~\mu$C/cm.  To obtain the total
specific heat, we added a constant term 60~J/mol K, describing
the background specific heat of a host lattice of heavy ions,
to the calculated from (\ref{specheat}) contribution of the hydrogen
subsystem to the specific heat.

\subsection{Non-zero field case}
\begin{figure}[hbt]
\begin{center}
\leavevmode
\epsfxsize=5.cm
\mframe{\rotate[r]{\epsffile{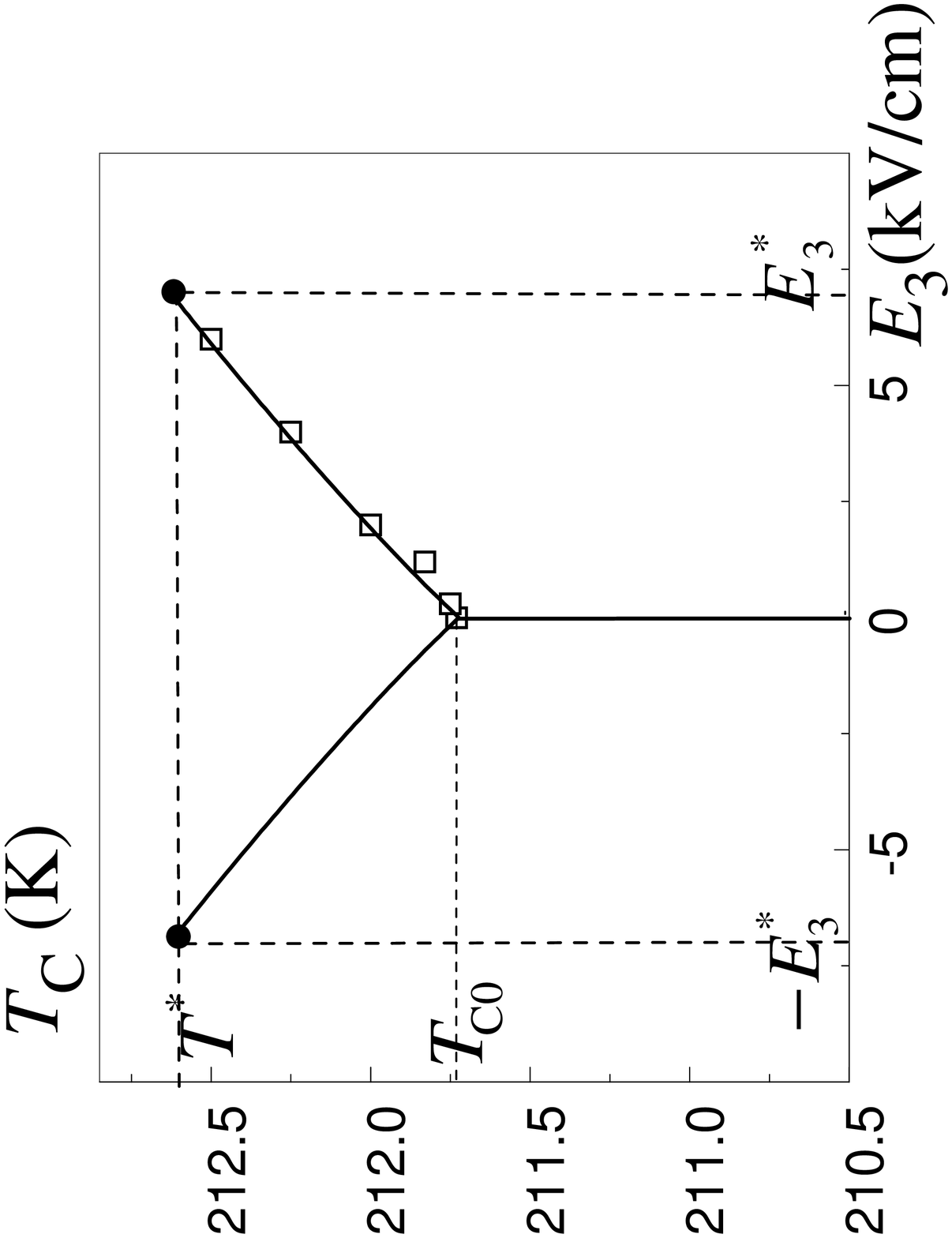}}}\hspace{0.5cm}
\epsfxsize=5.cm
\rotate[r]{\mframe{\epsffile{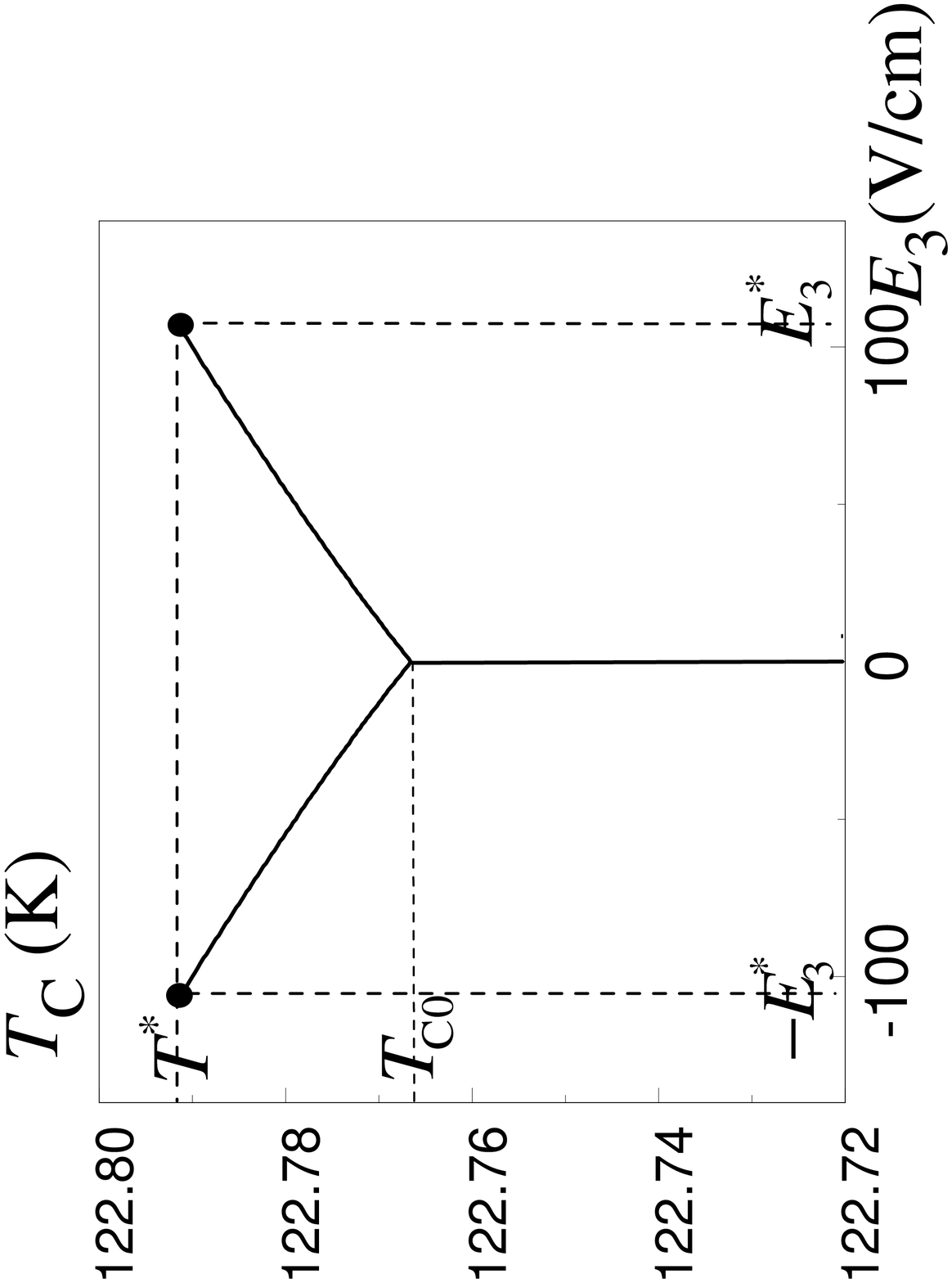}}}
\end{center}
\caption{\small
 $T_{\rm C}-E_3$  phase diagrams of KD$_2$PO$_4$ and \kdp.
Experimental points are taken from \protect\cite{Sidnenko}. } \label{pd}
\end{figure}

In figure~\ref{pd} we plot the $T_{\rm C}-E_3$ phase diagrams of
the \dekdp\ and \kdp\ crystals. These diagrams are of the same topology as
the $T_{\rm C}-\s$ diagram of  \dekdp\ \cite{our!!}.
The coordinates  of the critical point, calculated within the
microscopic theory, essentially depend on the magnitude of the order
parameter jump $\Delta\eta$ (the ratio of the polarization jump to the
saturation polarization) at the transition point
in zero field.  Since the order of the phase transition in a pure \kdp\
is close to the second one, while in a deuterated \dekdp\ a pronounced
jump of polarization  at the transition point takes place, the critical
field in \kdp\ is much lower than that in \dekdp.

For \dekdp\ we
obtain a good agreement with experimental data for a field dependence of
the transition temperature \cite{Sidnenko}:
$\partial T_{\rm C}/\partial E_3=0.13$K/ kVcm in our model and
$0.125\pm0.1$ in \cite{Sidnenko}.
Theoretical value of the critical field $E_3^*=7.0$~kV/cm, where the phase
transition in the  system disappears,  accords fairly well with
the experimental estimate $7.1\pm0.6$~kV/cm \cite{Sidnenko}. The critical
temperature $T^*=212.6$~K is close to that from the $T_{\rm C}-\s$ diagram.

For an undeuterated crystal, the theory
gives the following characteristics of the phase diagram

\begin{tabular}{ccc}
$E^*=105$~V/cm, & $T^*-T_0=0.05$~K, &   $\partial T_{\rm
C}/\partial E_3=0.23$K/ kVcm.
\end{tabular}

\noindent
The calculated critical field in \kdp\ agrees with the
the estimates $80\div370$~V/cm \cite{Schmidt}, obtained
via experimentally measured coefficients of Landau expansions. However,
at varying the microscopic $\Delta\eta$
within experimentally estimated limits (for which slight
changes in the theory parameters are required), the calculated critical
field is significantly increased, while more or less precisely measured
macroscopic characteristics (e.g. permittivity) remain
practically unchanged. We may conclude that the microscopic theory is able
only to {\it estimate} the coordinates of the critical point.

In addition, in fields close to critical, peaks of the longitudinal permittivity
or other characteristics having peculiarities at the transition point are
very sharp. Therefore, even small changes of temperature ($\sim10^{-3}$~K)
can change the permittivity value several times. The maximal
temperature step in calculations,
yielding more or less correct value of the permittivity
maximum, is $\sim10^{-4}$~K.

\begin{figure}[hbt]
\begin{center}
\leavevmode
\epsfxsize=6cm
\mframe{\rotate[r]{\epsffile{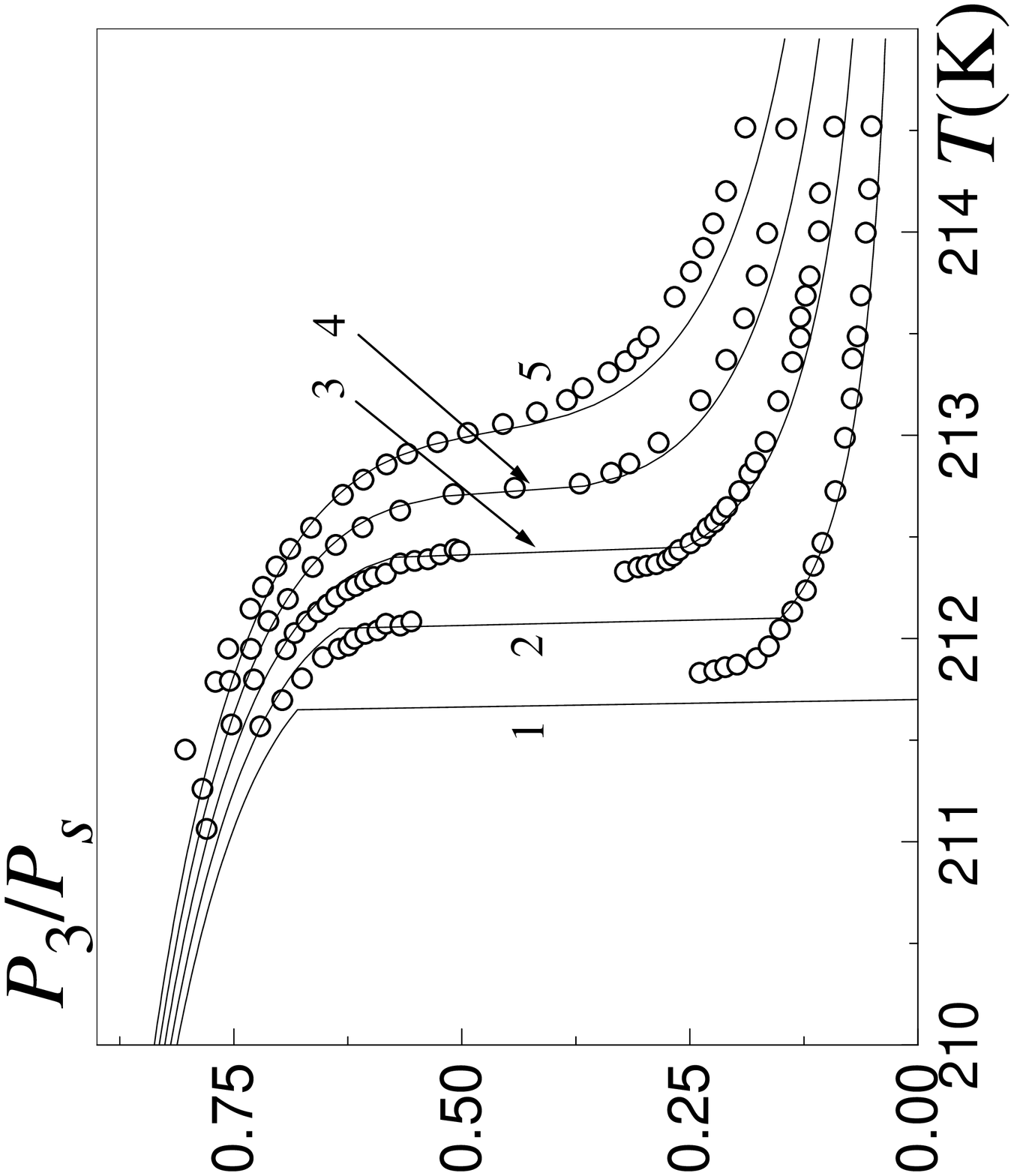}}}\hspace{0.5cm}
\epsfxsize=6cm
\mframe{\rotate[r]{\epsffile{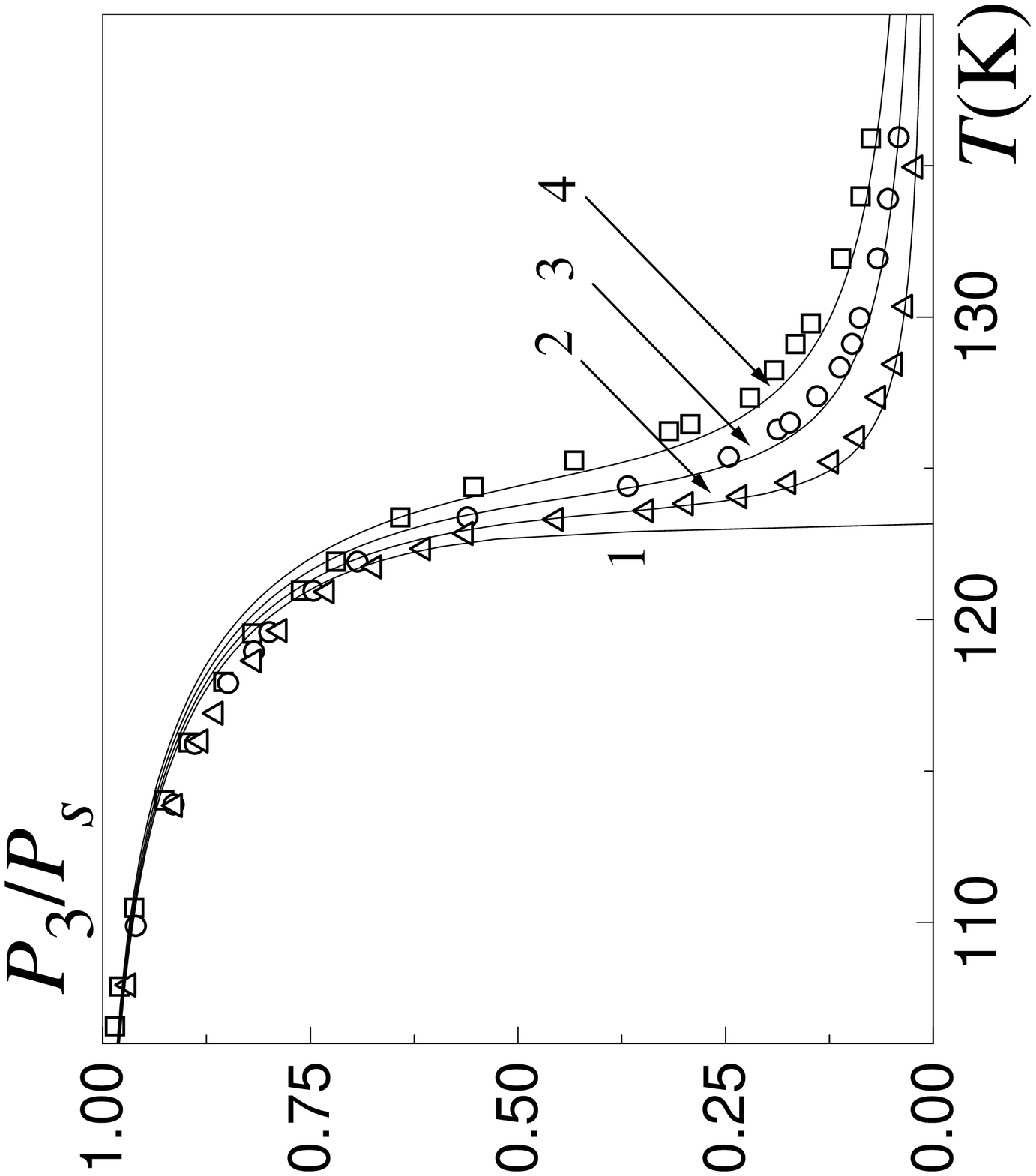}}}
\end{center}
\caption{\small
Ratio of polarization to saturation polarization
$P_3/P_s$ as a function of temperature for \dekdp\  (left) and
\kdp\ (right) at different values of electric field $E_3$ (kV/cm)
(left): 1 -- 0; 2 -- 2.82; 3 --  5.64;   4 -- 8.46; 5 -- 11.28;
(right): 1 -- 0; 2 -- 5.813; 3 -- 12.5 4 -- 20.31.
Experimental points are taken from
\protect\cite{Sidnenko2} for \kdp\ and from \protect\cite{Chabin} for \kdp.}
\label{pol} \end{figure}

As we have already mentioned, we cannot describe
polarization of the crystals, without introducing the ``ferroelectric''
values
of the dipole moment $\mu_3$. Nevertheless, we can compare the
field behavior of the order parameter $\eta$ with the
experimental data for the ratio $P_3/P_s$. Here $P_s=5.0~\mu$C/cm$^2$ for
\kdp\ and $6.2~\mu$C/cm$^2$ for \dekdp\ are the experimental
values of saturation polarization, practically independent of
external field \cite{Chabin}.  Such comparison is justified,
since the contributions to polarization from the terms $\chi_{33}^0E_3$ and
$e_{36}^0\e$ are not essential. Figure~\ref{pol} shows a fair accordance
of theoretical temperature and field dependences of $P_3/P_s$ for
\kdp\ and \dekdp\ with the experimental data. Increase of polarization and
vanishing of polarization jump are well reproduced by the theory.

\begin{figure}[hbt]
\begin{center}
\leavevmode
\epsfxsize=5.5cm
{\rotate[r]{\epsffile{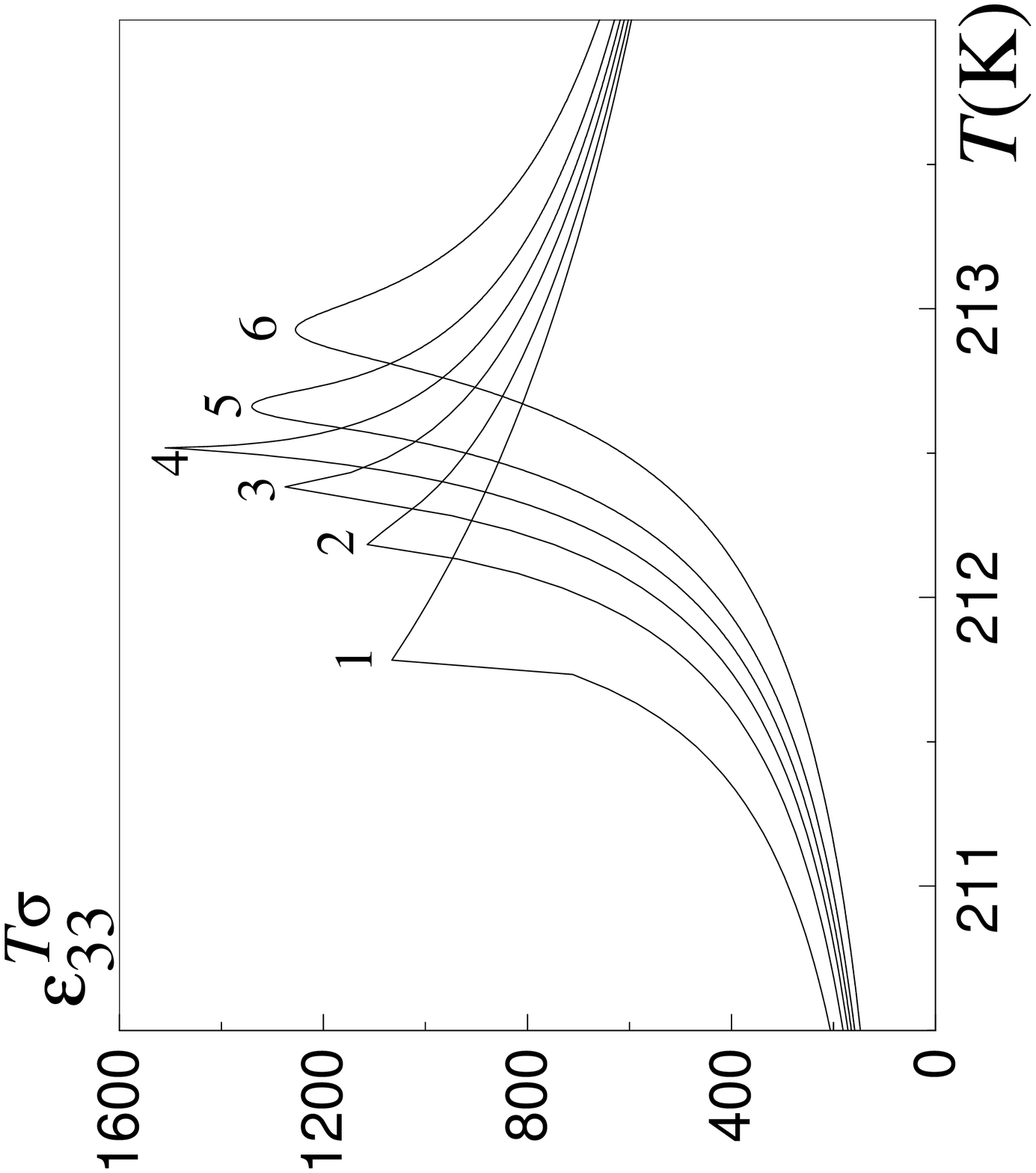}}}\hspace{0.5cm}
\epsfxsize=5.5cm
{\rotate[r]{\epsffile{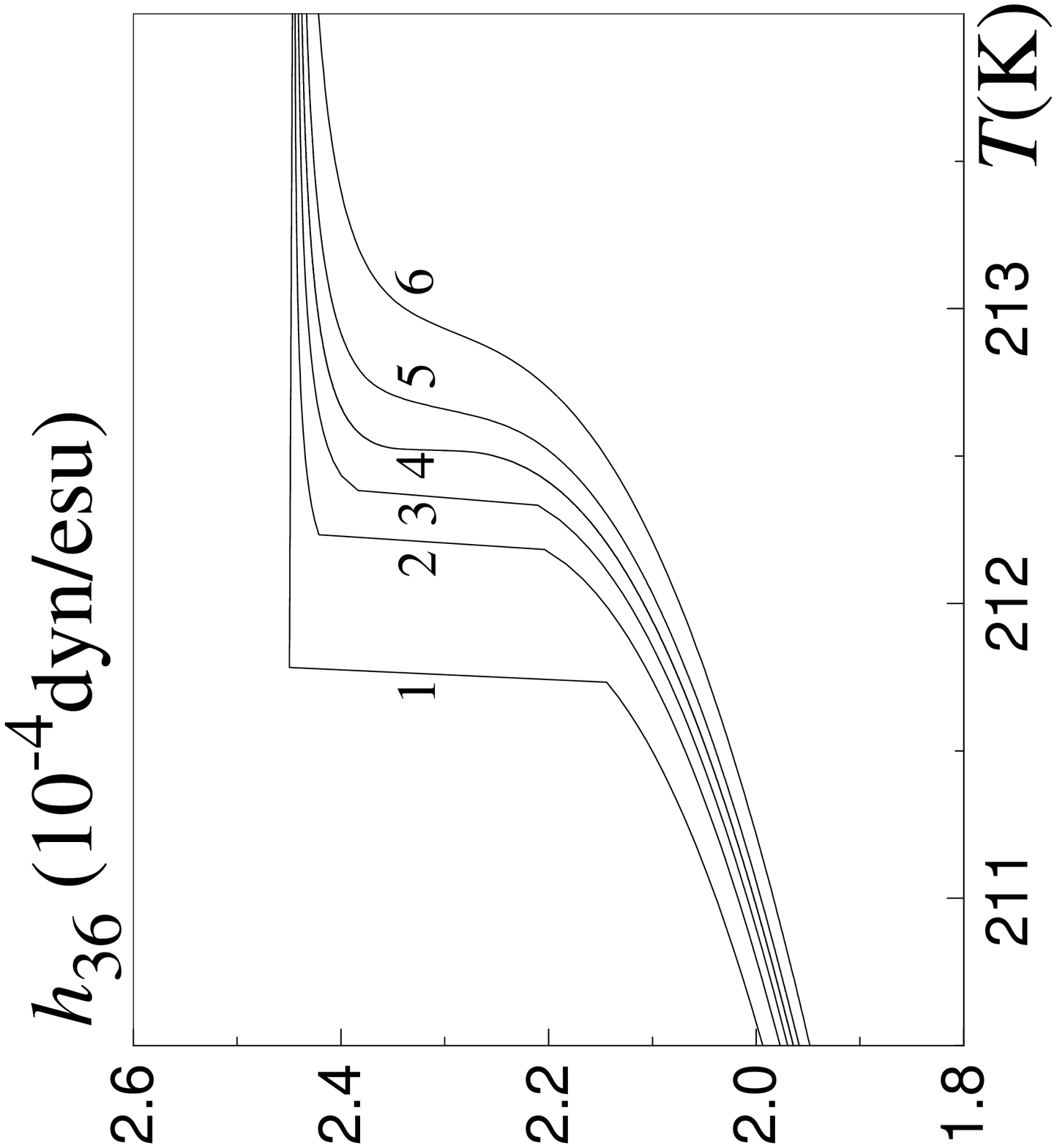}}}
\end{center}
\caption{\small
Isothermal static dielectric permittivity a free crystal
$\eps_{33}^{T\sigma}$ and isothermal piezomodule $h_{36}$ for a deuterated
 \dekdp\ as functions of temperature at different values of electric field
$E_3$ (kV/cm):
1 -- 0; 2
-- 3.82;
3 --  5.64;   4  -- $E^*=7.0$;  5 -- 8.46; 6 -- 11.28
.}
\label{hie} \end{figure}

In figure~\ref{hie} we depicted the temperature dependences of two
characteristics -- isothermal static dielectric permittivity of a free
crystal $\eps_{33}^{T\sigma}$ and the piezomodule $h_{36}$ of a deuterated
crystal at different values of external electric field. Temperature
curves of other characteristics of the crystal exhibiting
peculiarities at the transition point -- compliance $s_{66}^{E}$ and
piezomodules $d_{36}$ and $e_{36}$ -- are similar to those of the
permittivity $\eps_{33}^{T\sigma}$. Temperature and field behavior of the
piezomodule $g_{36}$ to those of $h_{36}$. The corresponding curves  for a
undeuterated are qualitatively similar, however, due to closeness of the
order of the phase transition to the second one, the temperature and
field intervals where all these changes take place are much more narrow.

Maximal values of $\eps_{33}^{T\sigma}$  and similar quantities shift
to higher temperatures with field. As the field approaches its critical
value, the peak values of these quantities in the transition points increase
and are maximal at the critical point. The higher fields smear out the phase
transition and round off and lower down the peaks.

The field effects on piezomodules $h_{36}$ and $g_{36}$ is similar
to its effects on polarization. On increasing the field, jumps of these
quantities at the transition point decrease and vanish at the critical
field. At higher fields, these quantities exhibit smooth temperature
dependences.

\begin{figure}[hbt]
\begin{center}
\leavevmode
\epsfxsize=5.5cm
\mframe{\rotate[r]{\epsffile{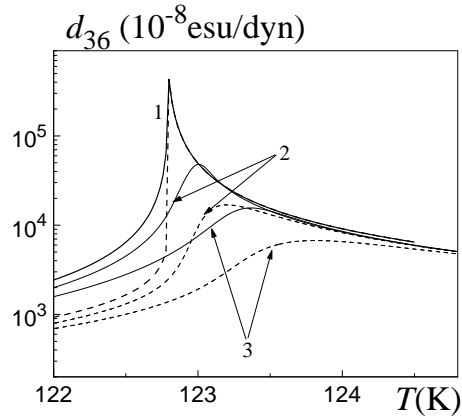}}}
\end{center}
\caption{\small
Isothermal and adiabatic piezomodules $d_{36}$ of a undeuterated \kdp\
as functions of temperature at different values of electric field $E_3$
(kV/cm):
1 -- 0;  2 -- 1; 3 --  2;   4  -- 3;  4 -- 5.}
\label{adia} \end{figure}

It is interesting to note that field dependences of adiabatic quantities
are stronger that those of the isothermal quantities.
Thus, as fig.~\ref{adia} where we present temperature curves of the adiabatic
and isothermal piezomodule $d^S_{36}$ of a pure \kdp\ shows, the peak values
of the adiabatic piezomodule (as well as of $s_{66}^{SE}$,
$e^S_{36}$, $\eps^S_{33}$) decrease with the external field (above the critical
point) much stronger than the peak values of the isothermal
$d^T_{36}$, and the maxima of the adiabatic quantities shift with
the field much more perceptibly that those of the isothermal quantities.

\begin{figure}[hbt]
\begin{center}
\leavevmode
\epsfxsize=5.5cm
\mframe{\rotate[r]{\epsffile{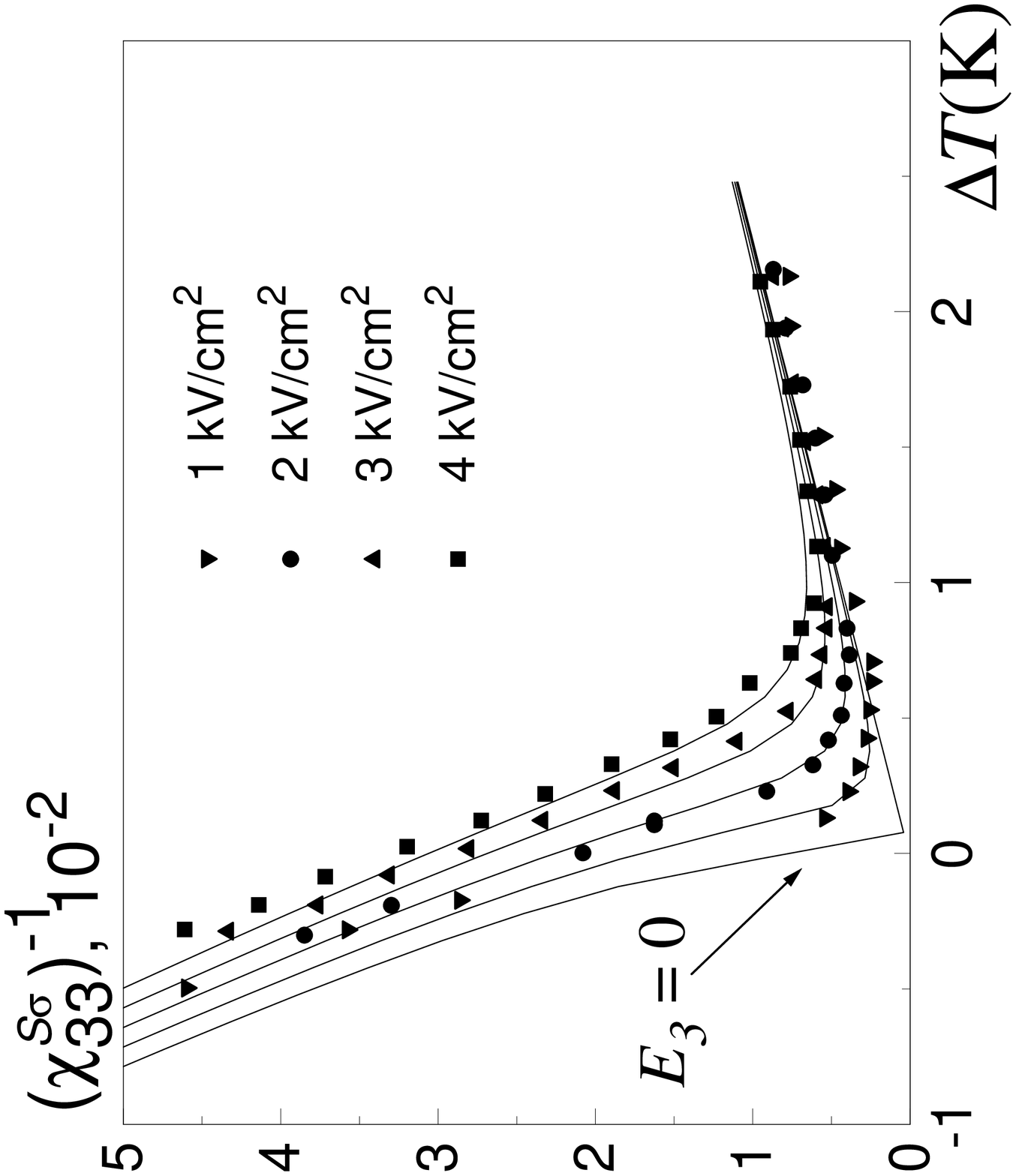}}}
\hspace{1em}
\epsfxsize=5.5cm
\mframe{\rotate[r]{\epsffile{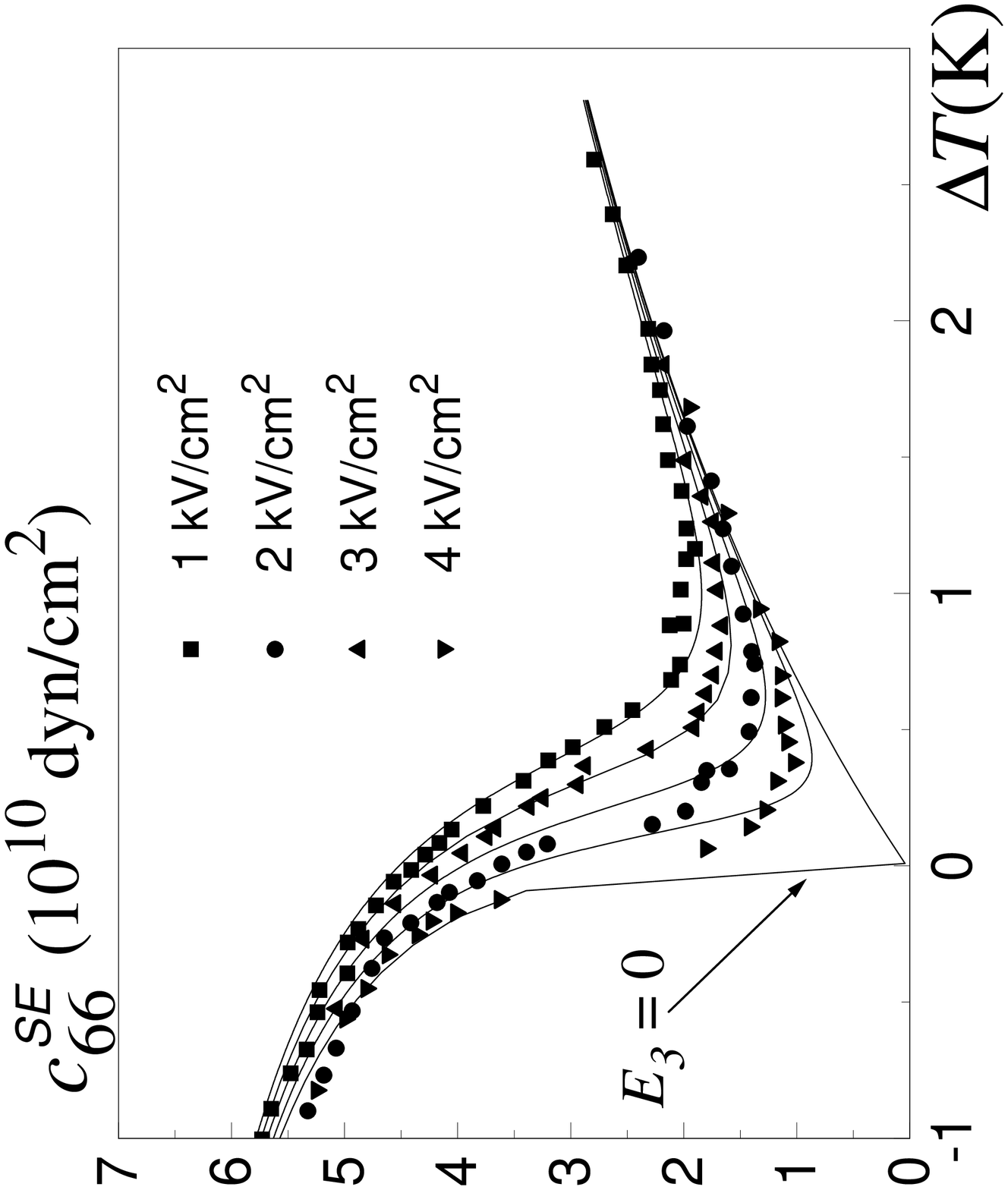}}}
\end{center}
\caption{\small
Adiabatic elastic constant  $c_{66}^{SE}$ and the inverse
longitudinal static dielectric
susceptibility of a free crystal $(\chi_{33}^{S\sigma})^{-1}$ for a pure
 \kdp\ as functions of temperature at different values of electric field $E_3$ (kV/cm).
Experimental points are taken from \protect\cite{Litov}.}
\label{litov} \end{figure}

As one can see (Fig.~\ref{litov}), the proposed theory  yields a fair
agreement with experiment for the field effects on the adiabatic
longitudinal dielectric susceptibility of  a free  crystal and
elastic constant $c_{66}^{SE}$ in a pure \kdp\ \cite{Litov}. Deviation of
the theoretical curves from experimental points in the vicinity of
the ``transition''
(maximum of the permittivity) is attributed to our neglecting tunneling
effects. Such neglecting worsens an agreement with an experiment for the
temperature curve of order parameter $P_3/P_s$ and, thereby,  of
$\chi_{33}^{S\sigma}$ and $c_{66}^{SE}$.

\section{Concluding remarks}
In this paper we apply the previously developed microscopic theory to description
of effects produced by the longitudinal
electric field  $E_3$ on the phase transition and physical properties of
 the \kdp\ family ferroelectrics.
Calculations are performed in the four-particle cluster approximation
within the proton ordering model without taking into account the tunneling
effects. The developed model takes into account
the existence of the shear strain $\e$, which is spontaneous in the
ordered phase and induced by the stress $\s$ and electric field $E_3$
(via the piezoeffect) in the disordered phase. We also take into account
other effects induced by this strain --
the splitting of short-range proton correlation energies and appearance of
effective molecular fields, created by piezoelectric coupling.

The presented model provides a fair agreement
with experimental data for temperature behavior  (at zero field)
of the dielectric, piezoelectric, elastic, and  thermal
characteristics -- second derivatives of thermodynamic potentials (dielectric
permittivity, piezomodules, elastic constants, specific heat) of the
crystals. Such agreement takes place for a deuterated crystal \dekdp\ and
for a pure \kdp\ without taking into account the tunneling.

The model satisfactorily reproduced the parameters of experimental
$T_{\rm C}-E_3$ phase diagrams of \dekdp\ and \kdp. The transition temperature
increases with field; this is accompanied by increase in polarization
in the high-temperature phase and decrease of its jump at the transition
point. The phase equilibrium curves terminate in the critical points, where
the polarization (order parameter) jump vanishes. At higher fields the temperature
curves of polarization and strain $\e$ are smooth. Such behavior is typical
for the systems undergoing the first order phase transitions in external fields
conjugate to the order parameter.

The magnitude of the critical field essentially depends on the degree
of the ``first-orderness'' of the phase transition).
Thus, in \kdp\ where the phase transition is close to the tricritical point,
the critical field is several times smaller than in \dekdp, where
a pronounced first order phase transition takes place.
Experimental method of the critical coordinates determination based in
the phenomenologic analysis allows only to estimate the critical field.
The calculated within the microscopic theory
value of the critical field is determined by the magnitude
of the order parameter at the transition point. Since
on changing this quantity (at the proper and slight change of the relations
between the theory parameters) within the limits of experimental estimates
the calculated critical field significantly increases,
the microscopic approach also permits only to estimate
the coordinates of the critical point, though this estimate is better than
the phenomenologic one.

As the field approaches the critical point, the peak values of thermodynamic
quantities that have peculiarities at the transition point (the longitudinal
dielectric permittivity  $\eps_{33}$, piezomodules $d_{36}$ and $e_{36}$, and
the elastic compliance  $s_{66}^E$) increase, while the jumps
of ``true'' constants of the crystals (piezomodules $g_{36}$ and $h_{36}$,
and elastic constant $c_{66}^P$) decrease. At the critical point
the peak values of the formers are maximal, whereas the jumps of the latters
vanish. In fields above the critical one, temperature dependences of all
thermodynamic quantities are smooth, and the maximal values of the characteristics
that have peculiarities at the transition point decrease.

In presence of external field, there arise a difference between
isothermal and adiabatic characteristics (dynamic methods of measurements
yield the adiabatic quantities), which have peculiarities
at the transition point. The field dependences of the
adiabatic quantities are stronger than those of the isothermal quantities.
Thus, in fields above the critical one, the peak values of the
adiabatic quantities decrease with the field much
faster that the peaks of the isothermal quantities, and
the maxima of the adiabatic quantities shift to higher temperatures that
the maxima of the isothermal quantities at the same field. Within the proposed
theory we obtain a fair description of the experimental data for the field
dependences of the adiabatic dielectric susceptibility and elastic constant
 of \kdp.

The developed model, as well all earlier used versions of proton
ordering model, requires two different values of effective dipole moment
to be introduced for description of polarization and dielectric
permittivity of the crystals (in paraelectric and ferroelectric phases).
For a consistent description of dielectric properties of these
crystals, phonon degrees of freedom and anharmonicity must be taken into
account.

\section*{Acknowledgments}
Authors would like to thank participants of the I Ukrainian-French Meeting
on  Ferroelectricity (Kiev, Ukraine, May 2000) for their interest to our
results.
This work was supported by the Foundation for
Fundamental Investigations of the Ukrainian Ministry in Affairs of Science
and Technology, project No 2.04/171.


\begin{thebibliography}{99}

\bibitem{Nel5}
R.O.~Piltz, M.I.~McMahon, and R.J.~Nelmes, Ferroelectrics  {\bf 108},
271 (1990).

\bibitem{our!}
I.V.Stasyuk,  R.R.Levitskii,   A.P.Moina.
// Phys. Rev. B., 1999, vol. 59, p. 8530-8540.



\bibitem{our!!}
I.V.Stasyuk, R.R.Levitskii, I.R.Zachek, A.P.Moina. // To be published in
Phys. Rev. B


\bibitem{stas1}
I.V. Stasyuk and I.N. Biletskii, Bull. Acad. Sci. USSR. Phys. Ser. {\bf 4},
79 (1983).



\bibitem{jps3}
I.V.Stasyuk,  R.R.Levitskii, I.R.Zachek,  A.P.Moina, A.S.Duda.  To appear
in Journ. of Phys. Studies.


\bibitem{Schmidt}
A.~Western, A.G.~Baker, C.P.~Bacon, V.H.~Schmidt.  Phys. Rev. B
{\bf 17}, 4461 (1978).

\bibitem{Litov}
E.Litov, C.W.Garland.  // Phys. Rev. B., 1970, vol.2, p. 4597-4602.

\bibitem{Chabin}
M. Chabin, F.Giletta.  // Ferroelectrics, 1977, vol. 15, p. 149-154.

\bibitem{Sidnenko}
V.V.~Gladkii and E.V.~Sidnenko. Sov. Phys. Solid State  {\bf 13}, 2592
(1972).

\bibitem{Sidnenko2}
 E.V.~Sidnenko and V.V.~Gladkii. Sov. Phys. Crystallogr.  {\bf 17}, 861
(1973).


\bibitem{Kobayashi}
J. Kobayashi, Y. Uesu, Y. Enomoto. // Phys. Status. Solidi B, 1971,
vol. 45,  p.293.

\bibitem{l366}
B.A.Strukov, M.Korzhuev, A. Baddur, V.A.Koptsik. //
Sov. Phys. Solid State,   1971, vol. 13, p.  1872-1877.

\bibitem{sugie}
K. Okada, H. Sugie. // Ferroelectrics, 1977, vol. 17, p. 325-327.


\bibitem{Schmidt19}
A.B.Western, A.G.~Baker, R.J. Pollina, V.H.Schmidt. // Ferroelectrics,
1977, vol. 17, p. 333.


\bibitem{Schmidt20}
V.H.Schmidt, A.B.Western, A.G.~Baker. // Phys.
Rev. Lett., 1976, vol. 37. p. 839-841.

\bibitem{Sidnenko3}
E.V.~Sidnenko and V.V.~Gladkii. // Sov. Phys. Crystallogr., 1973, vol. 18, p.
 138-142.

\bibitem{Vallade}
M. Vallade. // Phys. Rev.  B., 1975, vol. 12, p. 3755.

\bibitem{Blinc1966}
R. Blinc and S. Svetina, Phys. Rev. {\bf 147}, 430 (1966).

\bibitem{Blinc_Zeks}
R.~Blinc and B.~\v Zek\v s, {\it Soft modes in ferroelectrics and
antiferroelectrics} (Elseviers, New York, 1974).


\bibitem{Kanzig}
W.~ Kanzig, {\it Ferroelectrics and
antiferroelectrics} (Academic Press, New York, 1957).

\bibitem{a12}
Bantle W., Caflish C. Helv. Phys. Acta, 1943, vol. 16, p. 235.

\bibitem{a13}
Von Arx A., Bantle W. Helv. Phys. Acta, 1943, vol. 16, p. 211.

\bibitem{l147}
M. Lines, A. Glass. Ferroelectrics and related materials.


\bibitem{l370} F. Giletta, M. Chabin.  // Phys. Stat. Solidi B., 1980, vol. 100, p. K77-K82.

\bibitem{l379} A.S.Vasilevskaya, A.S.Sonin.//
// Sov. Phys. Solid State,  1971, vol. 13, p. 1550-1556.

\bibitem{l380}
B.A.Strukov, A. Baddur, V.A.Koptsik, I.A.Velichko. //
Sov. Phys. Solid State,   1972, vol. 14, p. 1034-1039.

\bibitem{a14}
Brody E.M., Cummins H.Z. Phys. Rev. Lett., 1968, vol. 21, p. 1263.

\bibitem{Mason_old}
W.P.Mason, Phys. Rev., 1946, vol. 69, p. 173

\bibitem{deQuirvane}
M. deQuervain, Helv. Phys.Acta {\bf 17} 509 (1944).




\bibitem{Zach1}
R.R.~Levitskii, I.R.~Zachek, and Ye.V.~Mits (unpublished).

\bibitem{Rhodes}
M. Tokunaga.  // J. Phys. Soc. Jap.,
1987, vol. 56, p. 1653-1656.

\bibitem{a15}
L.A.~Shuvalov, I.S.~Zheludev et al.
 Bull. Acad. Sci. USSR. Phys. Ser. {\bf 31},
1919 (1967).

\bibitem{Shuv}
L.A.~Shuvalov and A.V.~Mnatsakanyan, Sov. Phys. --  Crystallogr.  {\bf 11},
210 (1966).


\bibitem{l363} Wiseman G.G.
// Iee Transactions on electron devices, 1969, v.ed. 16, No6, p.588-593.


\bibitem{l365} Benepe J.W., W. Reese.  // Phys. Rev. B.,
1971, vol.3, p. 3032-3039.



\bibitem{l375} R.J.Mayer, J.L.Bjorkstam.  // J. Phys. Chem.
Solids, 1962, vol. 23, p. 619-620.


\bibitem{l384} J. Bornarel.  // Ferroelectrics, 1984, vol. 54, p. 245-248.

\bibitem{l412} W. Reese, L.F.May. // Phys. Rec. 1967, vol. 162, p  510-518.

\bibitem{l417} B.A. Strukov, M Amin, V.A. Koptsik.
// J. Phys. Soc. Japan, 1970, vol. 28, suppl. p. 207-209.


\end{thebibliography}
\end{document}